\documentclass[twocolumn,prb,showpacs]{revtex4}
\usepackage[ansinew]{inputenc}
\usepackage[dvips]{graphicx}
\usepackage{amsmath}
\usepackage{layout}
\usepackage{float}
\usepackage{subfigure}
\usepackage{amsfonts}
\usepackage{amssymb}

\begin{document}

\title{Axially symmetric focusing as a cuspoid diffraction catastrophe:\\Scalar and vector cases and comparison with the theory of Mie}
\begin{abstract}
An analytical description of arbitrary strongly aberrated axially symmetric
focusing is developed. This is done by matching the solution of geometrical
optics with a wave pattern which is universal for the underlying ray
structure. The corresponding canonical integral is the Bessoid integral, which
is a three-dimensional generalization of the Pearcey integral that
approximates the field near an arbitrary two-dimensional cusp. We first
develop the description for scalar fields and then generalize it to the vector
case. As a practical example the formalism is applied to the focusing of light
by transparent dielectric spheres with a few wavelengths in diameter. The
results demonstrate good agreement with the Mie theory down to Mie parameters
of about 30. Compact analytical expressions are derived for the intensity on
the axis and the position of the diffraction focus both for the general case
and for the focusing by microspheres. The high intensity region is narrower
than for an ideal lens of the same aperture at the expense of longitudinal
localization and has a polarization dependent fine structure, which can be
explained quantitatively. The results are relevant for aerosol and colloid
science where natural light focusing occurs and can be used in laser micro-
and nano-processing of materials.
\end{abstract}
\date{\today}

%

\author{Johannes Kofler}%

\email{johannes.kofler@univie.ac.at}\altaffiliation
{\\ This work was done at the Institute for Applied Physics, Johannes Kepler University Linz, Austria.}%

\affiliation
{Institute of Experimental Physics, University of Vienna, Boltzmanngasse 5, 1090 Vienna, Austria}

\author{Nikita Arnold}%

\email{nikita.arnold@jku.at}%

\affiliation
{Institute for Applied Physics, Johannes Kepler University Linz, Altenbergerstrasse 69, 4040 Linz, Austria}%


\maketitle

\section{Introduction}

Axially symmetric focusing of wave fields occurs in various areas of science,
since physical systems often possess an intrinsic rotational symmetry. In
particular, the electromagnetic field enhancement by small spherical particles
is important in many situations. Spheres have minimal surface energy for a
given volume and thus are naturally formed as a result of phase separation,
for example as aerosols or colloids. Applications of colloidal microspheres in
photonic crystals and photonic crystal slabs led to an explosion of the
experimental and theoretical studies of their optical
properties.\cite{Wij1998,Vla2001} The majority of these investigations
concentrate on their collective properties in a periodic arrangement. Single
microspheres are used as high quality optical resonators and as agents that
allow controlled and highly localized wavelength-dependent field enhancement
for non-linear optical studies and in resonance
spectroscopy.\cite{Gor1996,Vah2003} Here, the emphasis is placed on the
eigenmode analysis and the distribution of the field within the sphere or in
the immediate vicinity of its surface.

Lately, it was demonstrated that self-assembling arrays of transparent
colloidal microspheres can be employed for high-throughput laser-assisted
micro- and nano-structuring of materials.\cite{Bur1999,Bae2000,Bae2004}
Similar effects were observed in experiments on dry laser cleaning, where such
particles are used as controlled contaminants.\cite{Mue2001,Luk2002,Luk2003}
This necessitates better understanding of the focusing of light by
microspheres with diameters of several wavelengths. Only few rigorous results
are available for the intermediate range of sphere sizes and distances from
the particle. The majority of analytical approaches either deal with the
properties of the eigenmodes, or refer to the integral characteristics and/or
to the far field behavior.\cite{Boh1983} Mie resonances were analyzed on the
basis of advanced geometrical optics\cite{Rol2000} and detailed numerical
calculations for transparent spheres of several wavelengths in size were
performed in connection with the use of laser tweezers in
biology\cite{Roh2005} and for needs of aerosol science.\cite{Zim2001}

In this work we develop a theoretical description for an arbitrary
non-paraxial strongly aberrated axially symmetric focusing and apply it to the
case of dielectric microspheres. Our emphasis is on the fine structure of the
field distribution in the exterior of the sphere up to the focal region, which
can be used to control and improve the concentration of energy.

Strong spherical aberration makes the focusing non-trivial. Usually, the exact
solution is obtained using the Mie theory,\cite{Mie1908} which does not give
much of a physical insight as it requires the summation of a large number of
terms in a multipole expansion even for moderate sphere sizes. At the same
time, the main focusing properties of transparent dielectric microspheres
originate rather from the picture of geometrical optics.

One might think that in the lowest approximation a small sphere acts as an
ideal lens. However, in the range of sizes we are interested in, this picture
does not even provide a description which is qualitatively correct. Also
classical formulas for weak spherical aberration \cite{Bor2002} do not yield
useful results for the field behind a sphere: They predict that the maximum
intensity is kept unchanged and its position does not depend on the wavelength.

Our approach, following the method of uniform caustic
asymptotics,\cite{Kra1999} is based on the canonical integral for the cuspoid
ray topology of strong spherical aberration. Though this Bessoid integral ---
a member of the hierarchy of diffraction catastrophes\cite{Ber1980,Ber2001a}
--- appears naturally in the paraxial approximation, it can be used to
describe arbitrary axially symmetric strong spherical aberration by
appropriate coordinate and amplitude transformations. For angularly dependent
vectorial amplitudes the formalism uses higher-order Bessoid integrals.

The Bessoid integral is the axially symmetric generalization of the Pearcey
integral,\cite{Pea1946} which plays an important role in many short wavelength
phenomena.\cite{Con1981} Therefore, the present approach can be applied in
various areas of physics where axially symmetric focusing is of importance,
e.g., acoustics, semiclassical quantum mechanics, radio wave propagation and
scattering theory.

\section{The Bessoid integral}

\subsection{Definition}

We first consider the diffraction of a \textit{scalar} spherically aberrated
wave on a circular aperture with radius $a$ in the plane $z=-f$ around the
$z$-axis, where $f$ is the focal distance. The origin of the coordinate system
is put into the focus $F$. In cylindrical coordinates $(\rho,z)$, the paraxial
Fresnel--Kirchhoff diffraction integral \cite{Bor2002} yields the field
amplitude%
\begin{align}
U(\rho,z)  &  =-\dfrac{\text{i}\,k\,U_{0}}{f}\;\text{e}^{\,\text{i}%
\,k\,z}\nonumber\\
&  \;\;\times\int\nolimits_{0}^{a}J_{0}\!\left(  k\,\dfrac{\rho\,\widetilde
{\rho}_{1}}{f}\right)  \text{e}^{-\,\text{i}\,k\,\tfrac{z\,\widetilde{\rho
}_{1}^{2}}{2\,f^{2}}\,-\,\text{i}\,k\,B\,\widetilde{\rho}_{1}^{4}}%
\,\widetilde{\rho}_{1}\,\text{d}\widetilde{\rho}_{1}.
\label{eq Fresnel-Kirchhoff}%
\end{align}
Here $U_{0}$ is the amplitude of the incident wave in the center of the
aperture, $k$ is the wavenumber ($k=2\pi/\lambda$, where $\lambda$ is the
wavelength) and $\widetilde{\rho}_{1}$ is the distance from the axis on the
aperture. The Bessel function $J_{0}$ comes from the integration over the
polar angle $\varphi$. The parameter $B$ in the exponent determines the
strength of the spherical aberration. For $B>0$ the diffraction focus shifts
towards the aperture, while $B=0$ corresponds to ideal focusing.\cite{Bor2002}

\begin{figure}[t]
\begin{center}
\includegraphics{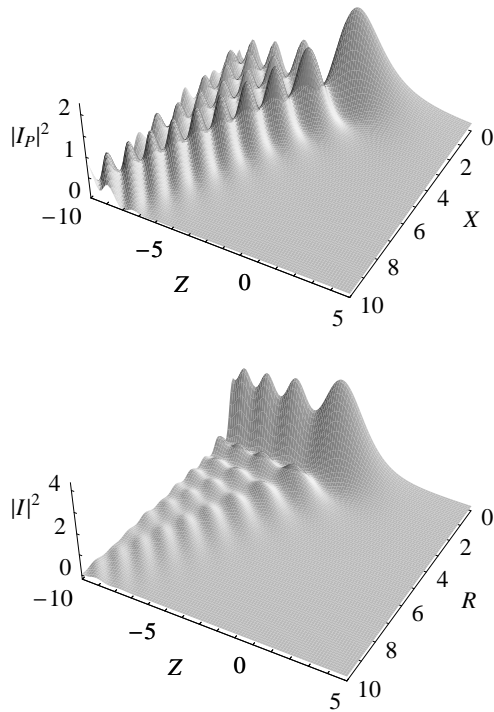}
\end{center}
\par
\vspace{-0.25cm} \caption{Absolute square of the Pearcey integral $I_{P}$
(top) and the Bessoid integral $I$ (bottom). The latter is proportional to the
field of a spherically aberrated wave within small angles approximation.}%
\label{fig Bessoid}%
\end{figure}

We introduce the dimensionless coordinates $\rho_{1}\equiv\sqrt[4]%
{4\,k\,B}\,\widetilde{\rho}_{1}$, $R\equiv\sqrt[4]{k^{3}/4\,B}\,\rho/f$ and
$Z\equiv\sqrt{k/4\,B}\,z/f^{2}$ and consider an infinitely large aperture.
Then the field (\ref{eq Fresnel-Kirchhoff}) becomes proportional to the
\textit{Bessoid integral} \cite{Kir2000}%
\begin{align}
I(R,Z)  &  =%
{\displaystyle\int\nolimits_{0}^{\infty}}
\rho_{1}\,J_{0}(R\,\rho_{1})\;\text{e}^{-\,\text{i}\,\left(  Z\,\tfrac
{\rho_{1}^{2}}{2}\,+\,\tfrac{\rho_{1}^{4}}{4}\right)  }\text{d}\rho
_{1}\label{eq Bessoid polar}\\
&  =\dfrac{1}{2\,\pi}\int\nolimits_{-\infty}^{\infty}\int\nolimits_{-\infty
}^{\infty}\text{e}^{\,\text{i}\,\phi}\,\text{d}x_{1}\text{d}y_{1},
\label{eq Bessoid}%
\end{align}
where%
\begin{equation}
\phi\equiv-R\,x_{1}-Z\,\dfrac{x_{1}^{2}+y_{1}^{2}}{2}-\dfrac{(x_{1}^{2}%
+y_{1}^{2})^{2}}{4}\,. \label{eq Phase Bessoid}%
\end{equation}
Its absolute square is shown in figure \ref{fig Bessoid}. In the Cartesian
representation $x_{1}=\rho_{1}\cos\varphi$ and $y_{1}=\rho_{1}\sin\varphi$ are
dimensionless coordinates in the plane of integration. Expression (\ref{eq
Bessoid}) is the axially symmetric generalization of the Pearcey integral
\cite{Pea1946}%
\begin{equation}
I_{P}(X,Z)=\dfrac{1}{\sqrt{2\,\pi}}\,%
{\displaystyle\int\nolimits_{-\infty}^{\infty}}
\,\text{e}^{-\,\text{i}\,\left(  X\,x_{1}\,+\,Z\,\tfrac{x_{1}^{2}}%
{2}\,+\,\tfrac{x_{1}^{4}}{4}\right)  }\text{d}x_{1},
\end{equation}
which is also shown in figure \ref{fig Bessoid}.

Both integrals correspond to so-called \textit{diffraction catastrophes}%
.\cite{Kra1999,Ber1980,Ber2001a} Their field distribution contains caustic
zones where the intensity predicted by geometrical optics goes to infinity.
The Pearcey integral corresponds to a cusp caustic, i.e., a single
one-dimensional curve in a two-dimensional space, and does not reveal a high
intensity along the axis, while the Bessoid integral corresponds to a
\textit{cuspoid caustic}, i.e., to a surface of revolution of the cusp in
three dimensions, as well as the caustic line up to the focus $F$ at $z=Z=0$.
The equation of the cusp is given by the semicubic parabola%
\begin{equation}
27\,R^{2}+4\,Z^{3}=0\,.
\end{equation}
Henceforth we will apply the term cusp also for the whole cuspoid. A caustic
is denoted as stable, if it does not change its topology under small
perturbations. This is the case for the Pearcey integral. The Bessoid integral
corresponds to a structurally unstable caustic, because an infinitely small
perturbation will destroy the radial symmetry and the axis will not be a
caustic zone any longer. It is, however, stable on the class of axially
symmetric wavefronts.

\begin{figure}[t]
\begin{center}
\includegraphics{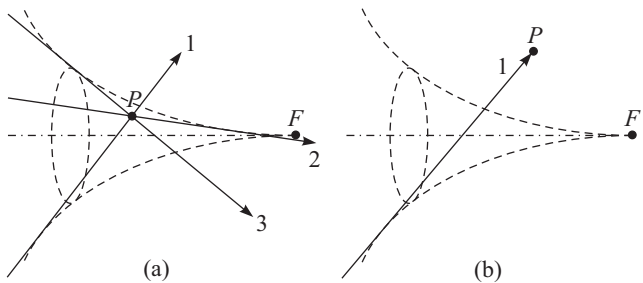}
\end{center}
\par
\vspace{-0.25cm} \caption{(a) 3-ray region inside the cuspoid (dashed line).
(b) 1-ray region outside. The $z$-axis is represented by a dashed-dotted
line.}%
\label{fig Rays}%
\end{figure}

The cusp is the envelope of the family of rays. The latter correspond to the
\textit{points of stationary phase} in the Bessoid integral, i.e., those
points where the two first partial derivatives with respect to $R$ and $Z$ of
the phase $\phi$ in (\ref{eq Bessoid}) vanish. Inside the cusp, for
$27\,R^{2}+4\,Z^{3}<0$, three rays (tangents to the cusp) arrive at each point
of observation $P\equiv(\rho,z)$, and outside, for $27\,R^{2}+4\,Z^{3}>0$,
there is only one real ray (figure \ref{fig Rays}). Thus, the cusp forms the
border between the lit region and the (partial) geometrical shadow, where two
rays merge.

Without loss of generality, we assume that all rays lie in the meridional
plane $\varphi=0$ ($y_{1}=0$) and hence correspond to the roots $x_{1,j}$
($j=1,2,3$) of the cubic equation%
\begin{equation}
R+Z\,x_{1}+x_{1}^{3}=0\,, \label{eq Cubic}%
\end{equation}
which are given by Cardan's formulas.\cite{Bro2004} On the axis, $R=0$, a cone
formed by an infinite number of rays converges. These rays originate from the
circle $x_{1}^{2}+y_{1}^{2}=-Z$ on the aperture. They are all in phase and
produce a high intensity along the axis (compare the two pictures in figure
\ref{fig Bessoid}). The oscillations occur due to interference with the ray
propagating along the $z$-axis. One can also directly observe that the Bessoid
integral has the topology of spherical aberration as the maximum of intensity
does not lie in the geometrical focus $Z=0$ but is spherically aberrated to a
negative value of $Z$, i.e., towards the aperture.

\subsection{Asymptotic expressions}

Off the caustic --- away from the cusp \textit{and the focal line} --- the
Bessoid integral (\ref{eq Bessoid}) can be approximated by the method of
stationary phase. As the integrand in (\ref{eq Bessoid}) is highly
oscillatory, the only significant contributions to the integral come from
those regions where the phase is stationary:\cite{Foc1956}%
\begin{equation}
I(R,Z)\approx%
{\displaystyle\sum\limits_{j=1}^{m}}
\,\dfrac{\text{e}^{\,\text{i}\,\phi_{j}\,+\,\text{i}\,\tfrac{\pi}%
{4}\,\text{sign}\,\mathbf{H}_{j}}}{\sqrt{\left|  \det\mathbf{H}_{j}\right|  }%
}\,, \label{eq Off caustic}%
\end{equation}
where the summation runs over all real rays, i.e., $m=1$ (lit region) or $m=3$
(shadow). The phase $\phi_{j}$ is obtained by inserting the $j$-th stationary
point $(x_{1,j},y_{1,j}\!=\!0)$ into the (\ref{eq Phase Bessoid}). The
determinant and signature of the Hessian are given by%
\begin{align}
\det\mathbf{H}_{j}  &  =Z^{2}+4\,x_{1,j}^{2}\,Z+3\,x_{1,j}^{4}%
\,,\label{eq Det H}\\
\text{sign}\,\mathbf{H}_{j}  &  =\text{sgn}(-Z-3\,x_{1,j}^{2})+\text{sgn}%
(-Z-x_{1,j}^{2})\,. \label{eq Sign H}%
\end{align}

Near the cusp the Bessoid integral shows an Airy-type behavior typical for
caustics where two real rays disappear and become complex.

One can also derive a different approximation valid \textit{on and near} the
caustic axis, i.e., for $Z\leq0$ and small $R$ (appendix \ref{Appendix Near
Axis Bessoid}):%
\begin{equation}
I(R,Z)\approx\dfrac{\sqrt{\pi}}{2}\,J_{0}(R\sqrt{-Z})\;\text{e}^{\,\text{i}%
\,\tfrac{Z^{2}-\pi}{4}}\,\operatorname{erf}\!\text{c\negthinspace}\left(
\dfrac{Z}{2}\;\text{e}^{\,\text{i}\,\tfrac{\pi}{4}}\right)
\text{\negthinspace}. \label{eq Near axis}%
\end{equation}
Here erfc is the complementary error function,\cite{Abr1993} which can also be
written in terms of Fresnel sine and cosine functions.\cite{Kir2000}
Expression (\ref{eq Near axis}) becomes exact at the axis $R=0$, where
$J_{0}(0)=1$. It shows that near the axis the Bessoid integral is virtually a
Bessel beam\cite{McG2005} with a variable cross section.

\subsection{Numerical evaluation}

As the Bessoid integrand is highly oscillatory, its evaluation for the whole
range of coordinates $R$ and $Z$ is non-trivial and of large practical
importance. Direct numerical integration along the real axis and the method of
steepest descent in the complex plane both have their disadvantages. By far
the fastest technique is based on the numerical solution of the
\textit{ordinary differential equation} (derivation in appendix \ref{Appendix
Differential})\cite{Kof2004,Time}%
\begin{equation}
L_{R}-Z\,I_{R}+\text{i}\,R\,I=0\,. \label{eq Ordinary}%
\end{equation}
Indices denote (partial) derivatives and $L\equiv I_{RR}+I_{R}/R$ is an
abbreviation for the radial Laplacian applied onto $I$. The three initial
conditions at $R=0$ are%
\begin{align}
I(0,Z)  &  =\dfrac{\sqrt{\pi}}{2}\;\text{e}^{\,\text{i}\,\tfrac{Z^{2}-\pi}{4}%
}\,\operatorname{erf}\!\text{c\negthinspace}\left(  \dfrac{Z}{2}%
\;\text{e}^{\,\text{i}\,\tfrac{\pi}{4}}\right)  \text{\negthinspace
},\label{eq Condition 1}\\
I_{R}(0,Z)  &  =0\,,\\
L(0,Z)  &  =Z\,I(0,Z)+\text{i}\,. \label{eq Condition 3}%
\end{align}
$I(0,Z)$ was taken from (\ref{eq Near axis}), $I_{R}(0,Z)$ vanishes due to
symmetry, and the last condition arises from the fact that the Bessoid
integral satisfies the paraxial Helmholtz equation $2\,$i$\,I_{Z}+L=0$, where
$I_{Z}$ is calculated from (\ref{eq Condition 1}).

In the literature the Pearcey integral was calculated by solving differential
equations,\cite{Con1984} by a series representation \cite{Con1973} and by the
first terms of its asymptotic expansion.\cite{Sta1983} The Bessoid integral
was expressed in terms of parabolic cylinder functions \cite{Jan1992} and as a
series.\cite{Kir2000} The latter work gives reference to an unpublished work
of Pearcey,\cite{Pea1963} stating that differential equations for the Bessoid
integral were employed there.

\subsection{Geometrical optics for the cuspoid}

In geometrical optics, the rays carry the information of amplitude and phase.
The total field in a point $P$ is given by the sum of all ray fields there. A
ray's field at $P$ is determined by\cite{Kra1990}%
\begin{equation}
U(P)=U_{0}\,\dfrac{\text{e}^{\,\text{i}\,k\,\psi}}{\sqrt{J}}\,, \label{eq Ray}%
\end{equation}
where $U_{0}$ is the amplitude at some initial wavefront, $\psi$ is the
eikonal, and $J$ is the generalized geometrical divergence, which can be
calculated from flux conservation along the ray. For a homogeneous medium with
constant refractive index\cite{Kra1990}%
\begin{equation}
J=\dfrac{R_{m}R_{s}}{R_{m0}R_{s0}}\,.
\end{equation}
$R_{m}$, $R_{s}$ are the main radii of curvature at the point $P$ and $R_{m0}%
$, $R_{s0}$ are the radii on the initial wavefront, where $U=U_{0}$.

When a ray touches a caustic, its radius of curvature (the geometrical
divergence in the general case) changes the sign and the ray undergoes a
\textit{caustic phase delay} \cite{Kra1990,Kra1999}\ of $-\pi/2$, which is
taken into account by the proper choice of the square root in (\ref{eq Ray}).
When a ray touches several caustics, these delays must be added. The total
caustic phase shift, denoted as $\Delta\varphi$, can be explicitly written in
the phase. For the cuspoid topology and ray numbering ($j=1,2,3$) according to
figure \ref{fig Rays}, we obtain:%
\begin{equation}
U(P)=U_{0}\,\dfrac{\text{e}^{\,\text{i}\,k\,\psi}}{\sqrt{J}}=U_{0}%
\,\dfrac{\text{e}^{\,\text{i}\,k\,\psi\,+\,\text{i}\,\Delta\varphi}}%
{\sqrt{\left|  J\right|  }} \label{eq Caustic phase shift}%
\end{equation}
with%
\begin{equation}
\Delta\varphi_{j}=\left\{
\begin{array}
[c]{rl}%
-\pi & \text{ for }j=1,\\
0 & \text{ for }j=2,\\
-\pi/2 & \text{ for }j=3.
\end{array}
\right.
\end{equation}
Ray 1 touched the cuspoid and the focal line, ray 2 is not shifted, and ray 3
touched the cuspoid.

\section{Relation between geometrical and wave optics}

\subsection{Matching with the Bessoid integral}

If we have found the phases $\varphi\equiv k\,\psi$ and divergences $J$ of the
rays, the (scalar) geometrical optics solution with an axially symmetric 3-ray
cuspoid topology can be written as%
\begin{equation}
U(\mathbf{r})=\sum_{j=1}^{3}\dfrac{U_{0,j}\;\text{e}^{\,\text{i}\,\varphi
_{j}(\mathbf{r})}}{\sqrt{J_{j}(\mathbf{r})}}\,.
\label{eq Geometrical solution}%
\end{equation}
Here $\mathbf{r}\equiv(\rho,z)$ are the real-space coordinates and we have
allowed for different initial amplitudes $U_{0,j}$ of the rays. This field
shows singularities at the caustic, especially on the axis, which is the most
interesting region for applications.

We want to describe \textit{arbitrary axially symmetric focusing} by matching
the solution of geometrical optics (where it is correct) with a wave field
constructed from the Bessoid integral (\ref{eq Bessoid}), which naturally
appears in the paraxial approximation and is finite everywhere, and its
partial derivatives $I_{R}$ and $I_{Z}$ (\textit{method of uniform caustic
asymptotics}). We make the \textit{Ansatz} \cite{Kra1999}%
\begin{equation}
U=\left(  A\,I+\dfrac{1}{\text{i}}\,A_{R}\,I_{R}+\dfrac{1}{\text{i}}%
\,A_{Z}\,I_{Z}\right)  \text{e}^{\,\text{i}\,\chi}\,. \label{eq Ansatz}%
\end{equation}
The yet unknown arguments of the Bessoid integral and its derivatives are
$\mathbf{R}\equiv(R(\mathbf{r}),Z(\mathbf{r}))$. $A(\mathbf{r})$,
$A_{R}(\mathbf{r})$ and $A_{Z}(\mathbf{r})$ are three amplitude factors and
$\chi(\mathbf{r})$ is a phase function. (The indices $R$ and $Z$ in the
amplitudes do not indicate derivatives.) Now the geometrical optics solution
(\ref{eq Geometrical solution}) is matched with the stationary phase
approximation of (\ref{eq Ansatz}) by equating the amplitudes and
phases:\cite{Kra1999,Bre1992}%
\begin{align}
\dfrac{U_{0,j}}{\sqrt{J_{j}}}  &  =\dfrac{A(\mathbf{r})+A_{R}(\mathbf{r}%
)\,\phi_{R}(\mathbf{R},\mathbf{t}_{j})+A_{Z}(\mathbf{r})\,\phi_{Z}%
(\mathbf{R},\mathbf{t}_{j})}{\sqrt{H_{j}}}\,,\label{eq Matching amplitude}\\
\varphi_{j}(\mathbf{r})  &  =\chi(\mathbf{r})+\phi(\mathbf{R},\mathbf{t}%
_{j})\,. \label{eq Matching phase}%
\end{align}
$\phi_{R}$ and $\phi_{Z}$ are the partial derivatives of (\ref{eq Phase
Bessoid}) and%
\begin{equation}
\dfrac{1}{\sqrt{H_{j}}}\equiv\dfrac{\text{e}^{\,\text{i}\,\tfrac{\pi}%
{4}\,\text{sign}\,\mathbf{H}_{j}}}{\sqrt{\left|  \det\mathbf{H}_{j}\right|  }%
}\,, \label{eq Hessian lit}%
\end{equation}
where the determinant and signature of the Hessian are written in (\ref{eq Det
H}) and (\ref{eq Sign H}), respectively. Outside the cusp, the rays 2 and 3
are complex and the general definition of $H_{j}$ is more subtle, namely
\begin{equation}
\dfrac{1}{\sqrt{H_{j}}}\equiv\sqrt{\dfrac{\text{i}}{\phi_{20,j}}}%
\,\sqrt{\dfrac{\text{i}}{\phi_{02,j}}}\,, \label{eq Hessian general}%
\end{equation}
with $\phi_{20}\equiv\partial^{2}\phi$/$\partial x_{1}^{2}$, $\phi_{02}%
\equiv\partial^{2}\phi/\partial y_{1}^{2}$ (the index $j$ denotes substitution
of the $j$-th point of stationary phase as argument).

The three points of stationary phase were denoted as $\mathbf{t}_{j}%
\equiv(t_{j},0)$, where the $t_{j}$ are given by the (correctly ordered)
Cardan's solutions of (\ref{eq Cubic}), i.e., of%
\begin{equation}
R+Z\,t+t^{3}=0\,. \label{eq Stationary points}%
\end{equation}
Note that they are functions of the Bessoid coordinates, $\mathbf{t}%
_{j}=\mathbf{t}_{j}(\mathbf{R})$, and the latter depend on the real space
coordinates: $\mathbf{R}=\mathbf{R}(\mathbf{r})$. The partial derivatives with
respect to $R$ and $Z$ in (\ref{eq Matching amplitude}) must be evaluated in
such a way as the $t_{j}$ were held constant, although they are functions of
$\mathbf{R}$ themselves. The conditions (\ref{eq Matching amplitude}) and
(\ref{eq Matching phase}) give 6 equations for the 6 unknowns $R$, $Z$, $\chi
$, $A$, $A_{R}$, and $A_{Z}$.

It is convenient to solve (\ref{eq Matching phase}), that is%
\begin{equation}
\varphi_{j}=\chi-R\,t_{j}-\dfrac{1}{2}\,Z\,t_{j}^{2}-\dfrac{1}{4}\,t_{j}%
^{4}\,,
\end{equation}
using quantities that are permutationally invariant with respect to the roots
$t_{j}$.\cite{Con1981,Bre1992} This yields%
\begin{align}
R  &  =\sqrt{\dfrac{Z^{3}}{54}-\dfrac{4\,b_{2}}{9\,Z}}\,,\nonumber\\
Z  &  =\pm\sqrt[4]{\dfrac{2}{3}}\,\sqrt{-2\,\text{sgn}(b_{3})\,\sqrt{b_{2}%
+q}+2\,\sqrt{D}}\,,\label{eq R, Z, chi}\\
D  &  \equiv2\,b_{2}-q+2\sqrt{b_{2}^{2}-b_{2}\,q+q^{2}}\,,\nonumber\\
\chi &  =b_{1}-\dfrac{1}{6}\,Z^{2}\,,\nonumber
\end{align}
where sgn$(Z)=\;$sgn$(Z^{4}-24\,b_{2})$. The $b_{l}$ ($l=1,2,3$) are given by
$b_{1}\equiv(1/3)$\thinspace$%
{\textstyle\sum\nolimits_{j=1}^{3}}
\varphi_{j}$, $b_{2}\equiv%
{\textstyle\sum\nolimits_{j=1}^{3}}
(\varphi_{j}-b_{1})^{2}$ and $b_{3}\equiv%
{\textstyle\sum\nolimits_{j=1}^{3}}
(\varphi_{j}-b_{1})^{3}$. The quantity $q$ (sometimes called discriminant) can
be expressed in different ways:%
\begin{align}
q^{3}  &  \equiv6\,b_{3}^{2}-b_{2}^{3}=\dfrac{1}{2^{11}}\,R^{2}\,(27\,R^{2}%
+4\,Z^{3})^{3}\nonumber\\
&  =-2\,(\varphi_{1}-\varphi_{2})^{2}\,(\varphi_{2}-\varphi_{3})^{2}%
\,(\varphi_{3}-\varphi_{1})^{2}\,.
\end{align}
Hence, it vanishes exactly at the caustic where two phases are equal. At the
cuspoid $\varphi_{2}=\varphi_{3}$ ($27\,R^{2}+4\,Z^{3}=0$) and on the axis
$\varphi_{1}=\varphi_{3}$ ($R=0$).

The solutions of (\ref{eq Matching amplitude}), that is%
\begin{equation}
\dfrac{U_{0,j}}{\sqrt{J_{j}}}=\dfrac{A-t_{j}\,A_{R}-\frac{1}{2}\,t_{j}%
^{2}\,A_{Z}}{\sqrt{H_{j}}}\,, \label{eq Matching amplitudes}%
\end{equation}
are\cite{Bre1992}%
\begin{align}
A  &  =-\,U_{0,1}\,\dfrac{\sqrt{H_{1}}}{\sqrt{J_{1}}}\,\dfrac{t_{2}\,t_{3}%
}{(t_{3}-t_{1})\,(t_{1}-t_{2})}-...-...\,,\nonumber\\
A_{R}  &  =U_{0,1}\,\dfrac{\sqrt{H_{1}}}{\sqrt{J_{1}}}\,\dfrac{t_{1}}%
{(t_{3}-t_{1})\,(t_{1}-t_{2})}+...+...\,,\label{eq A, AR, AZ}\\
A_{Z}  &  =2\,U_{0,1}\,\dfrac{\sqrt{H_{1}}}{\sqrt{J_{1}}}\,\dfrac{1}%
{(t_{3}-t_{1})\,(t_{1}-t_{2})}+...+...\,,\nonumber
\end{align}
where the cyclic terms permutate the numbering of rays: $(1,2,3)\rightarrow
(2,3,1)\rightarrow(3,1,2)$. The Bessoid matching solution (\ref{eq Ansatz})
does not show the divergences of geometrical optics.

Note that this method utilizes also the so-called complex rays which have less
apparent physical meaning. It turns out that both real and complex rays
provide the \textit{geometrical skeleton for the wave flesh}.\cite{Kra1999}

\subsection{Expressions on and near the axis in the general case}

All formulas can be strongly simplified on and near the axis inside the
cuspoid (small $\rho$, $z<0$). The Bessoid coordinates have the simple form
(appendix \ref{Appendix Near axis coordinates})%
\begin{align}
R  &  \approx\dfrac{(\varphi_{1}-\varphi_{3})/2}{\sqrt{2}\,\sqrt[4]%
{(\varphi_{1}+\varphi_{3})/2-\varphi_{2}}}\approx\dfrac{k\,\rho\,\sin\beta
}{\sqrt{-Z}}\,,\label{eq Near axis R}\\
Z  &  \approx-2\,\sqrt{\dfrac{\varphi_{1}+\varphi_{3}}{2}-\varphi_{2}}%
\approx-2\,\sqrt{\varphi_{\text{np}}-\varphi_{\text{p}}}\,,
\label{eq Near axis Z}%
\end{align}
where $\beta>0$ is the local angle of the non-paraxial cone of rays with the
axis and $\varphi_{\text{np}}$ and $\varphi_{\text{p}}$ denote the phases of
the non-paraxial rays and the (par)axial ray, respectively (see figure
\ref{fig Nearaxis} in appendix \ref{Appendix Near axis coordinates}). The
simple natural combination%
\begin{equation}
R\sqrt{-Z}\approx\dfrac{\varphi_{1}-\varphi_{3}}{2}\approx k\,\rho\,\sin
\beta\label{eq Bessel argument}%
\end{equation}
also appears in the near axis approximation for the Bessoid integral (\ref{eq
Near axis}). On the axis ($\rho=0$, $\varphi_{1}=\varphi_{3}$) we obtain $R=0$
and $Z=-2\,\sqrt{\varphi_{1}-\varphi_{2}}$.

The results (\ref{eq Near axis R})--(\ref{eq Bessel argument}) have
transparent physical meaning. Indeed, near the axis the largest contribution
to the field comes from the converging cone of non-paraxial rays (similar to
ray 1) that intersect the axis at an angle $\beta$. If the angle $\beta$ is
constant and all rays have the same intensity, the result is the Bessel
beam.\cite{McG2005} Such beams have a propagation constant along the
$z$-direction equal to the $z$-component of the wavevector of the plane waves
which form them and correspondingly the argument of the Bessel function
(cylindrical analog of a plane wave) is equal to $k\,\rho\,\sin\beta$. As the
angle $\beta$ gradually changes for the spherically aberrated wave, so does
the argument of the Bessel function.

Additionally, there exists the axial ray 2, which is not present in the
canonical Bessel beam (though it often appears in real experimental
situations). The interference of this beam with the converging ray cone
results in the intensity oscillations along the axis (figure \ref{fig
Bessoid}, bottom). Clearly, these oscillations are largely due to the phase
difference $\varphi_{\text{np}}-\varphi_{\text{p}}$. At large negative $Z$ in
(\ref{eq Near axis}) erfc$[\tfrac{Z}{2}\exp($i$\tfrac{\pi}{4})]\rightarrow
2\,[1\!-\!\tfrac{1}{\sqrt{\pi}\,Z}\exp($i$\tfrac{3\pi-Z^{2}}{4})]$, and the
oscillating behavior is governed by the phase of the exponent, which is equal
to $3\,\pi/4-(\varphi_{\text{np}}-\varphi_{\text{p}})$. This clarifies the
origin of expression (\ref{eq Near axis Z}), as it is $Z^{2}$ entering the
final formulas.

In particular, the global maximum is expected on axis at the first
constructive interference of the axial and the non-paraxial rays. Because
$Z<0$ in this region, the two terms of the erfc expansion are first in phase
when the phase difference is $\varphi_{1}-\varphi_{2}=3\,\pi/4$. The
geometrical meaning of this result is that rays 1 and 3 are shifted by
$-\pi/2$ as they touch the cusp. In addition, they acquire a further shift of
$-\pi/2$ when crossing the focal line. But exactly \textit{on the axis} only
half of this delay has occurred yet, which yields the $3\,\pi/4$ difference.
The numerical maximum of the Bessoid intensity (absolute square) occurs at
$Z_{\text{m}}\approx-3.051$ and hence this yields the condition%
\begin{equation}
\varphi_{1}-\varphi_{2}=\dfrac{Z_{\text{m}}^{2}}{4}\approx2.327\,,
\label{eq Phase difference axis}%
\end{equation}
which is close to $3\,\pi/4\approx2.356$.

The width of the focal line caustic, $\rho_{w}$, is defined by the first zero
$w_{0}\approx2.405$ of the Bessel function in (\ref{eq Near axis}). Hence,
with (\ref{eq Bessel argument}),%
\begin{equation}
\rho_{w}\equiv\dfrac{w_{0}}{k\,\sin\beta}\approx0.383\,\dfrac{\lambda}%
{\sin\beta}\,. \label{eq Caustic width}%
\end{equation}
In the geometrical optics picture the first minimum occurs when rays 1 and 3
interfere destructively, i.e., when their phase difference becomes $\pi$. This
results in $\varphi_{1}-\varphi_{3}=\pi+\pi/2$, where the term $\pi/2$ takes
into account the caustic phase shift of ray 1: $\rho_{w}\approx(\varphi
_{1}-\varphi_{3})/2\,k\,\sin\beta=0.375\,\lambda/\sin\beta$. Note, that this
is smaller than the Airy spot for the same aperture angle\cite{Bor2002} and
large angles $\beta$ are indeed realized, e.g., in the case of the sphere
studied below.

Finally, we present an expression for the field (\ref{eq Ansatz}) on the axis.
The equations for the amplitudes (\ref{eq A, AR, AZ}) simplify tremendously
(appendix \ref{Appendix On axis}) and result in%
\begin{equation}
U=\left[  \,\dfrac{U_{0,1}\,\sqrt{2\,k\,\rho\,\sin\beta}}{\sqrt{J_{1}}}\left(
\text{i}\,I-\dfrac{1}{Z}\right)  +\dfrac{U_{0,2}}{\sqrt{J_{2}}}\,\right]
\text{e}^{\,\text{i}\,\varphi_{2}}\,. \label{eq On axis field}%
\end{equation}
The structure of expression (\ref{eq On axis field}) helps to understand its
physical meaning. It details the contribution of the cone of non-paraxial
rays, represented by ray 1 (first term), and the axial ray 2 (second term) to
the overall structure of the field. Note that in the general case not only the
angle $\beta$, but also the amplitude of the converging cone may vary along
$z$ ($Z$), thus slowly modifying the properties of the Bessel beam in the
axial region. This enters (\ref{eq On axis field}) via amplitude
transformations and is manifested by the presence of the initial ray
amplitudes in both terms. Inside the cusp on the axis ($z,Z<0$ and
$\rho,R\rightarrow0$) both $1/\sqrt{J_{2}}$ and the ratio $\sqrt{\rho}%
/\sqrt{J_{1}}$ remain finite as the divergence of the paraxial ray 2 is
non-singular, while the sagittal divergence $J_{1}$ of the cone of
non-paraxial rays 1 is proportional to $\rho$. Due to the Bessoid matching
procedure the singularity of the converging cone is removed by the
compensating factor $\sqrt{\rho}$. Along the axis the last term in (\ref{eq On
axis field}) partly cancels with the second term in the parentheses of the
first term. As a result, the on axis field behavior up to the focus is
dominated by a single term proportional to the Bessoid integral $I$, which
justifies the maximum condition (\ref{eq Phase difference axis}) discussed above.

\subsection{Angular dependences and vectorial problems: Higher-order Bessoid matching}

Often --- especially in \textit{vectorial} problems --- there exists axial
symmetry with respect to the wavefronts, ray phases and generalized
divergences, but not with respect to the amplitudes. In this case, new
functions are required to represent arbitrary angular dependence of the field.
The natural generalization of (\ref{eq Bessoid}) are the \textit{higher-order
Bessoid integrals}\cite{Jan1992} with the non-negative integer $m$:%
\begin{equation}
I_{m}(R,Z)=%
{\displaystyle\int\nolimits_{0}^{\infty}}
\rho_{1}^{m+1}\,J_{m}(R\,\rho_{1})\;\text{e}^{-\,\text{i}\,\left(
Z\,\tfrac{\rho_{1}^{2}}{2}\,+\,\tfrac{\rho_{1}^{4}}{4}\right)  }\text{d}%
\rho_{1}, \label{eq Higher-order Bessoids}%
\end{equation}
where $I_{0}\equiv I$ and $J_{m}$ are higher-order Bessel functions. The
higher-order Bessoid integrals obey the recurrence relation%
\begin{equation}
I_{m+1}=-I_{m,R}+m\,\dfrac{I_{m}}{R}\,. \label{eq Bessoid recursive}%
\end{equation}
The integral $I_{m}$ is canonical for angular dependent geometrical field
components $U^{(m)}(\rho,z)\,\sin m\,\varphi$ or $U^{(m)}(\rho,z)\,\cos
m\,\varphi$. In matching similar to (\ref{eq Ansatz}),%
\begin{equation}
U^{(m)}=\left(  A_{m}\,I_{m}+\dfrac{1}{\text{i}}\,A_{mR}\,I_{m,R_{m}}%
+\dfrac{1}{\text{i}}\,A_{mZ}\,I_{m,Z_{m}}\right)  \text{e}^{\,\text{i}%
\,\chi_{m}}\!, \label{eq Ansatz higher-order}%
\end{equation}
the angular dependence cancels. Here $A_{m}$, $A_{mR}$ and $A_{mZ}$ are the
higher-order amplitude factors, whereas $I_{m,R_{m}}$ and $I_{m,Z_{m}}$ are
partial derivatives of the higher-order Bessoid integrals $I_{m}$. Since the
latter can be written in terms of $I_{0}$, it can be shown that the points of
stationary phase, the matching of phases and thus the higher-order coordinates
($R_{m}$, $Z_{m}$) and phases ($\chi_{m}$) are identical with the original
ones:%
\begin{equation}
R_{m}=R\,,\;\;Z_{m}=Z\,,\;\;\chi_{m}=\chi\,.
\end{equation}
From the physical point of view, this reflects the \textit{conservation of the
wavefront} and thus the ray phases and divergences.

The equations for the amplitudes have to be generalized. The higher-order
amplitudes $A_{m}$, $A_{mR}$ and $A_{mZ}$ have the same form as (\ref{eq A,
AR, AZ}), but with an additional factor $($i$\,t_{j})^{m}$ in each
denominator, i.e.,%
\begin{align}
A_{m}  &  =-\,U_{0,1}\,\dfrac{\sqrt{H_{1}}}{\sqrt{J_{1}}}\,\dfrac{t_{2}%
\,t_{3}}{(\text{i}\,t_{1})^{m}\,(t_{3}-t_{1})\,(t_{1}-t_{2})}%
-...-...\,,\nonumber\\
A_{mR}  &  =U_{0,1}\,\dfrac{\sqrt{H_{1}}}{\sqrt{J_{1}}}\,\dfrac{t_{1}%
}{(\text{i}\,t_{1})^{m}\,(t_{3}-t_{1})\,(t_{1}-t_{2})}+...+...\,,\nonumber\\
A_{mZ}  &  =2\,U_{0,1}\,\dfrac{\sqrt{H_{1}}}{\sqrt{J_{1}}}\,\dfrac
{1}{(\text{i}\,t_{1})^{m}\,(t_{3}-t_{1})\,(t_{1}-t_{2})}+...+...\,.
\label{eq Am}%
\end{align}
A more detailed description of the higher-order Bessoid integrals as well as
the derivation of the recurrence relation and the amplitude equations can be
found in appendix \ref{Appendix Higher-order}.

\section{The sphere}

\subsection{Geometrical optics solution}

\begin{figure}[t]
\begin{center}
\includegraphics{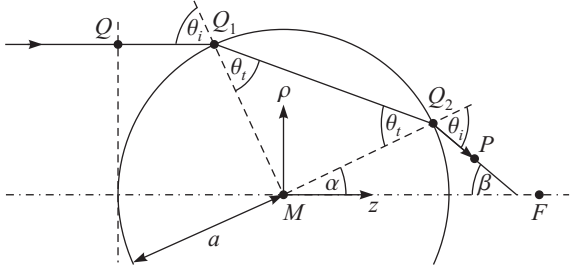}
\end{center}
\caption{Refraction of a ray --- propagating from $Q$ to $P$ --- by a sphere
with radius $a$ and refractive index $n$. The picture is drawn in the
meridional plane and all indicated angles are positive.}%
\label{fig Sphere}%
\end{figure}

Consider a plane wave falling on a transparent sphere in vacuum. Figure
\ref{fig Sphere} illustrates the refraction of a single ray in the meridional
plane, containing the point of observation $P$ and the axis. Within the frame
of geometrical optics the cuspoid is formed behind the sphere in analogy to
figure \ref{fig Rays}.

Let $a$ be the sphere radius and $n>1$ its refractive index. In contrast to
the previous sections, we choose the origin of the axially symmetric
cylindrical coordinate system $(\rho,z)$ differently now, namely as the center
$M$ of the sphere. The incident plane wave propagates parallel to the
$z$-axis. The geometrical optics focus, formed by the paraxial rays, is
located at $F\equiv(0,f)$ with \cite{Ber1978}%
\begin{equation}
f\equiv\dfrac{a}{2}\,\dfrac{n}{n-1}\,.
\end{equation}
A ray passes the point $Q$, is first refracted at $Q_{1}$, a second time at
$Q_{2}$ and propagates to $P$. The incident and transmitted angle, $\theta
_{i}$ and $\theta_{t}$, are related by Snell's law, $\sin\theta_{i}%
=n\sin\theta_{t}$. Writing the position of $P\equiv(\rho,z)$ in polar
coordinates, $\rho=l\sin\theta$ and $z=l\cos\theta$, one can find the
following expression, determining the three rays that arrive at $P$:%
\begin{equation}
l\sin(\theta+2\,\theta_{i}-2\,\theta_{t})=a\sin\theta_{i}\,,
\label{eq Cubic like sphere}%
\end{equation}
where one has to substitute $\theta_{t}=\arcsin[(\sin\theta_{i})/n]$. This is
a transcendental cubic-like equation which has three roots, either all real or
one real and two complex conjugate. (For $n\geq\sqrt{2}$ this is true for
$z\geq a$; if $n<\sqrt{2}$, the 3-ray region does not start until some
distance behind the sphere.) We denote them as $\theta_{i,j}(j=1,2,3)$ and
choose their order consistently with the previous notations. Therefore,
$\theta_{i,1}$ is always real and negative, whereas $\theta_{i,2}$ and
$\theta_{i,3}$ are either real and positive (lit region) with $\theta
_{i,2}<\theta_{i,3}$ or complex conjugate (geometrical shadow).

When the $\theta_{i,j}$ are known, we find the $\theta_{t,j}$ from Snell's law
and the $\alpha_{j}$ and $\beta_{j}$ from%
\begin{equation}
\alpha=2\,\theta_{t}-\theta_{i}\,,\quad\beta=2\,\theta_{i}-2\,\theta_{t}\,.
\label{eq alpha beta}%
\end{equation}
Omitting the index $j$, the three ray coordinates can be written as
\begin{equation}
s\equiv\overline{Q_{2}P}=\dfrac{l\cos\theta-a\cos\alpha}{\cos\beta}\,.
\label{eq Path}%
\end{equation}
The eikonal is the optical path accumulated from $Q$ to $P$ (on the dashed
vertical line in figure \ref{fig Sphere} all rays are still in phase):%
\begin{align}
\psi &  =\overline{QQ_{1}}+n\,\overline{Q_{1}Q_{2}}+\overline{Q_{2}%
P}-a\nonumber\\
&  =a\,(2\,n\cos\theta_{t}-\cos\theta_{i})+s\,. \label{eq Eikonal sphere}%
\end{align}
The sphere radius $a$ was subtracted from the path contributions to make the
eikonal zero in the center $M$, if there were no sphere.

Next we calculate the geometrical optics amplitudes by determining the
meridional and sagittal radii of curvature, $R_{m}$ and $R_{s}$, and their
changes due to refraction. Formulas for the refraction on an arbitrary surface
with arbitrary orientation of the main radii exist in the
literature.\cite{Kra1990,Cer2001} A simple derivation for the sphere can be
found in appendix \ref{Appendix Radii}. It yields the dependence of the actual
radii of curvature $R_{m}$ and $R_{s}$ (right after the refraction) on the
initial radii $R_{m0}$ and $R_{s0}$ (just before the refraction):%
\begin{align}
R_{m}  &  =\dfrac{n\,a\,R_{m0}\cos^{2}\theta_{t}}{a\cos^{2}\theta_{i}%
+R_{m0}\,(\cos\theta_{i}-n\cos\theta_{t})}\,,\label{eq Rm}\\
R_{s}  &  =\dfrac{n\,a\,R_{s0}}{a+R_{s0}\,(\cos\theta_{i}-n\cos\theta_{t})}\,.
\label{eq Rs}%
\end{align}
For a plane wave, $R_{m0},R_{s0}\rightarrow\infty$, the radii of curvature in
the points $Q_{1}$ (inside the sphere) and $Q_{2}$ (outside the sphere) have
the compact form%
\begin{align}
R_{m,Q_{1}}  &  =-a\,\dfrac{\sin\theta_{i}\cos^{2}\theta_{t}}{\sin(\theta
_{i}-\theta_{t})}\,,\label{eq R1}\\
R_{m,Q_{2}}  &  =-a\,\dfrac{\cos\theta_{i}}{2}\left(  \dfrac{\cos\theta
_{i}\sin\theta_{t}}{\sin(\theta_{i}-\theta_{t})}-1\right)  \!,\\
R_{s,Q_{1}}  &  =-a\,\dfrac{\sin\theta_{i}}{\sin(\theta_{i}-\theta_{t})}\,,\\
R_{s,Q_{2}}  &  =-a\,\dfrac{\sin(2\,\theta_{t}-\theta_{i})}{\sin(2\,\theta
_{i}-2\,\theta_{t})}\,. \label{eq R4}%
\end{align}
The overall geometrical generalized divergence after both refractions reads
(index $j$ omitted)%
\begin{align}
\dfrac{1}{\sqrt{J}}  &  =\dfrac{\sqrt{R_{m,Q_{1}}\,R_{s,Q_{1}}}}%
{\sqrt{(R_{m,Q_{1}}+d)\,(R_{s,Q_{1}}+d)}}\nonumber\\
&  \quad\quad\quad\times\dfrac{\sqrt{R_{m,Q_{2}}\,R_{s,Q_{2}}}}{\sqrt
{(R_{m,Q_{2}}+s)\,(R_{s,Q_{2}}+s)}}\,, \label{eq Divergence sphere}%
\end{align}
where $d\equiv2\,a\cos\theta_{t}$ is the distance of propagation within the
sphere. Note that ray 1 has a negative angle $\theta_{i}$. Besides, a double
caustic phase shift should be added (manually) to the phase of this ray (minus
sign) as in (\ref{eq Caustic phase shift}). The caustic shifts of the rays 2
and 3 are taken into account automatically if the branch cut for the square
roots in (\ref{eq Divergence sphere}) is along the negative real axis from
$-\infty$ to $0$ and the branch with $\sqrt{-1}=+$i is used. In this procedure
it is not allowed to multiply the radicands and write them under one common
square root. Also the case of complex rays 2 and 3 is covered correctly by
this convention.

Finally, the geometrical optics solution for the sphere is given by (\ref{eq
Geometrical solution}), where the eikonal $\psi$ and divergence $J$ are given
by (\ref{eq Eikonal sphere}) and (\ref{eq Divergence sphere}). The equation
determining the three rays is (\ref{eq Cubic like sphere}). In the geometrical
shadow the sum (\ref{eq Geometrical solution}) becomes only the term with
$j=1$.

\begin{figure}[t]
\begin{center}
\includegraphics{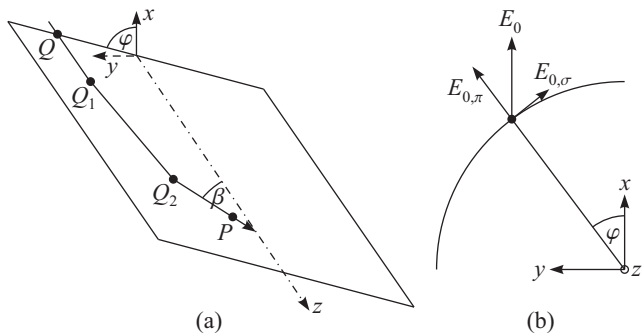}
\end{center}
\caption{(a) A ray propagates from $Q$ to $P$ in the meridional plane (plane
of incidence). (b) Decomposition of the initial electric field vector with
length $E_{0}$ into its $\pi$- and $\sigma$-component parallel and
perpendicular to the meridional plane}%
\label{fig Projection}%
\end{figure}

To incorporate Fresnel transmission coefficients, we assume that the incident
light is linearly polarized in $x$-direction, i.e., the incident electric
field vector is%
\begin{equation}
\mathbf{E}_{0}=E_{0}\,\mathbf{e}_{x}\,,
\end{equation}
with $\mathbf{e}_{x}$ the unit vector in $x$-direction and $E_{0}\equiv U_{0}%
$. Since axial symmetry is broken, we introduce the polar angle $\varphi$
which is measured from $x$ to $y$. The point of observation $P\equiv
(\rho,\varphi,z)$ will be reached by three rays (two may be complex) and their
angles $\theta_{i,j}$ are still determined by (\ref{eq Cubic like sphere}),
for all three rays lie in the meridional plane, containing $P$ and the
$z$-axis (figure \ref{fig Projection}a). The initial $\pi$- and $\sigma
$-polarized components depend on $\varphi$ (figure \ref{fig Projection}b):
\begin{align}
E_{0,\pi}  &  =E_{0}\cos\varphi\,,\\
E_{0,\sigma}  &  =E_{0}\sin\varphi\,.
\end{align}
We define the overall transmission coefficients
\begin{align}
T_{\pi}  &  \equiv t_{12,\pi}\,t_{21,\pi}=1-r_{12,\pi}^{2}\,,\\
T_{\sigma}  &  \equiv t_{12,\sigma}\,t_{21,\sigma}=1-r_{12,\sigma}^{2}\,.
\end{align}
Here the $t_{12}$ ($r_{12}$) are the standard Fresnel transmission
(reflection) coefficients \cite{Bor2002} from the medium 1, i.e., vacuum, into
the medium 2, i.e., the sphere.

The ray field behind the sphere is found by the projection onto the original
Cartesian system $(x,y,z)$. We write the components of the transmission vector
$\mathbf{T}\equiv(T_{x},T_{y},T_{z})$ and show the ray index $j=1,2,3$
explicitly. The $\varphi$-dependence is indicated with the superscript $(m)$:%
\begin{equation}%
\begin{array}
[c]{lll}%
T_{x,j}=T_{j}^{(0)}+T_{j}^{(2)}\cos2\,\varphi\,, &  & T_{j}^{(0)}\equiv
\dfrac{T_{\pi,j}\cos\beta_{j}+T_{\sigma,j}}{2}\,,\smallskip\\
T_{y,j}=T_{j}^{(2)}\sin2\,\varphi\,, &  & T_{j}^{(1)}\equiv T_{\pi,j}\sin
\beta_{j}\,,\smallskip\\
T_{z,j}=T_{j}^{(1)}\cos\varphi\,, &  & T_{j}^{(2)}\equiv\dfrac{T_{\pi,j}%
\cos\beta_{j}-T_{\sigma,j}}{2}\,.
\end{array}
\label{eq Transmission components}%
\end{equation}
Hence, the geometrical optics solution for the electric field $\mathbf{E}%
\equiv(E_{x},E_{y},E_{z})$ --- including the eikonal $\psi$ (\ref{eq Eikonal
sphere}) and divergence $J$ (\ref{eq Divergence sphere}) --- reads%
\begin{equation}%
\begin{array}
[c]{lll}%
E_{x,j}=E_{j}^{(0)}+E_{j}^{(2)}\cos2\,\varphi\,, &  & \\
E_{y,j}=E_{j}^{(2)}\sin2\,\varphi\,, &  & E_{j}^{(m)}\equiv E_{0}%
\,\dfrac{T_{j}^{(m)}\,\text{e}^{\,\text{i}\,k\,\psi_{j}}}{\sqrt{J_{j}}}\,.\\
E_{z,j}=E_{j}^{(1)}\cos\varphi\,, &  &
\end{array}
\label{eq Electric field components}%
\end{equation}

\subsection{The Bessoid matching solution}

\begin{figure}[t]
\begin{center}
\includegraphics{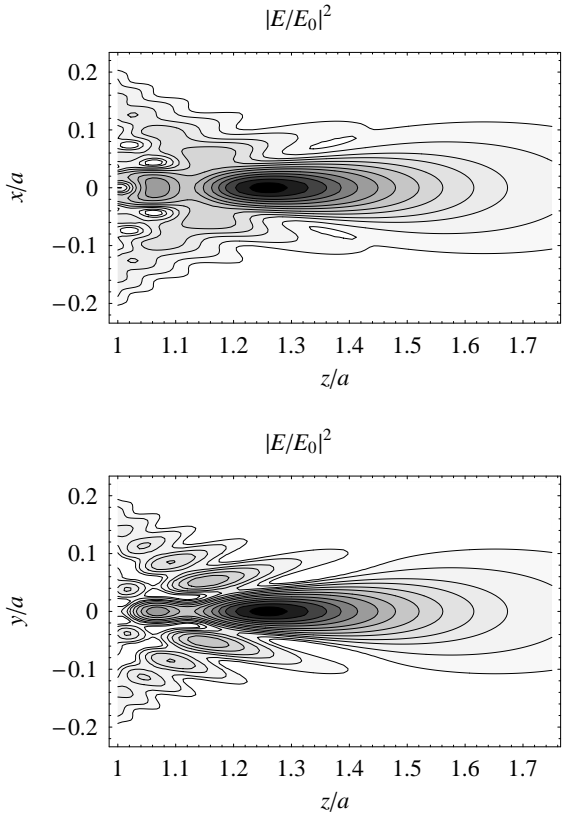}
\end{center}
\par
\vspace{-0.25cm} \caption{Normalized intensity $\left|  E/E_{0}\right|  ^{2}$
in the normalized $x$,$z$-plane (top) and in the $y$,$z$-plane (bottom).
Contour shadings go from white (zero) to black ($\approx700$). Parameters:
refractive index $n=1.5$, dimensionless wavenumber $k\,a=100$. The initial
electric field vector is $\mathbf{E}_{0}=E_{0}\,\mathbf{e}_{x}$. The sphere
with radius $a$ is situated in the origin. The focus of geometrical optics is
located at $z=f=1.5\,a$, whereas the diffraction focus (the point of maximum
intensity) is significantly shifted towards the sphere: $f_{d}\approx1.25\,a$.
In dimensional units, for a wavelength of $\lambda=0.248\,$$\mu$m the sphere
radius is $a\approx4\,$$\mu$m.}%
\label{fig Contourxzyz}%
\end{figure}

Matching each term $E^{(m)}=\sum\nolimits_{j=1}^{3}E_{j}^{(m)}$ by the Ansatz
(\ref{eq Ansatz}) in its higher-order formulation (\ref{eq Ansatz
higher-order}) with the corresponding integral $I_{m}$, we obtain the
vectorial electric field $\mathbf{E}\equiv(E_{x},E_{y},E_{z})$ in the form%
\begin{equation}
\mathbf{E}=E^{(0)}\!\left(
\begin{array}
[c]{c}%
1\\
0\\
0
\end{array}
\right)  +E^{(1)}\!\left(
\begin{array}
[c]{c}%
0\\
0\\
\cos\varphi
\end{array}
\right)  +E^{(2)}\!\left(
\begin{array}
[c]{c}%
\cos2\,\varphi\\
\sin2\,\varphi\\
0
\end{array}
\right)  \!. \label{eq Electric field matched}%
\end{equation}
Figure \ref{fig Contourxzyz} illustrates the intensity, i.e., the absolute
square of the electric field $\left|  E\right|  ^{2}\equiv\mathbf{E}%
\,\mathbf{E}^{\ast}$, for $\varphi=0$ ($x$,$z$-plane) and $\varphi=\pi/2$
($y$,$z$-plane).

The magnetic field $\mathbf{H}$ can be calculated similarly (incident magnetic
field $\mathbf{H}_{0}=H_{0}\,\mathbf{e}_{y}$, $H_{0}=E_{0}$) and the
(normalized) Poynting vector is given by $\mathbf{S\equiv}\operatorname{Re}%
(\mathbf{E\!}\times\mathbf{\!H}^{\ast})$.

\subsection{On the axis}

On the axis the electric field is given by its $x$-component only (direction
of polarization) due to averaging over $\varphi$ in (\ref{eq Electric field
matched}). For $z<f$ (inside the cusp) it is given by the analytical
expression (\ref{eq On axis field}). After several simplifications
\cite{Kof2004} it can be written as%
\begin{equation}
E=E_{0}\left[  \,T_{1}\,D_{1}\!\left(  \text{i}\,I-\dfrac{1}{Z}\right)
+\dfrac{T_{2}}{1-z/f}\,\right]  \text{e}^{\,\text{i}\,\varphi_{2}}\,,
\label{eq Electric field on axis}%
\end{equation}
where the transmission factors $T_{j}\equiv T_{j}^{(0)}$ are given in (\ref{eq
Transmission components}) and for dielectric spheres have the form:%
\begin{align}
T_{1}  &  =\dfrac{n\,(1+3\cos\beta_{1})\cos\theta_{i,1}\cos\theta_{t,1}%
}{(n\cos\theta_{i,1}+\cos\theta_{t,1})^{2}}\,,\\
T_{2}  &  =\dfrac{4\,n}{1+n^{2}}\,.
\end{align}
The phases in the coordinate $Z=-2\,\sqrt{\varphi_{1}-\varphi_{2}}$ are%
\begin{align}
\varphi_{1}  &  =\varphi_{3}=k\,a\left(  2\,n\cos\theta_{t,1}-\cos\theta
_{i,1}+\dfrac{\sin\alpha_{1}}{\sin\beta_{1}}\right)
\!,\label{eq Phases sphere 1}\\
\varphi_{2}  &  =2\,k\,a\,(n-1)+k\,z\,, \label{eq Phases sphere 2}%
\end{align}
and $D_{1}\equiv\sqrt{\varphi_{1}-\varphi_{3}}/\sqrt{J_{1}}$ is the first
ray's \textit{compensated sagittal divergence}:%
\begin{align}
D_{1}  &  =-\dfrac{\sqrt{(R_{m,Q_{1}})_{1}\,(R_{s,Q_{1}})_{1}}}{\sqrt
{[(R_{m,Q_{1}})_{1}+d_{1}]\,[(R_{s,Q_{1}})_{1}+d_{1}]}}\nonumber\\
&  \quad\quad\quad\times\dfrac{\sqrt{(R_{m,Q_{2}})_{1}\,(R_{s,Q_{2}})_{1}}%
}{\sqrt{(R_{m,Q_{2}})_{1}+s_{1}}}\,\sqrt{2\,k}\,\sin\beta\nonumber\\
&  =-2\,\sqrt{2\,k\,a}\,\cos(\beta_{1}/2)\nonumber\\
&  \quad\quad\quad\times\sqrt{\dfrac{\cot\theta_{i,1}\cos\theta_{t,1}%
\sin(\beta_{1}/2)}{1+1/n^{2}-3\sin^{2}(\beta_{1}/2)/\sin^{2}\theta_{i,1}}}\,,
\label{eq Compensated divergence}%
\end{align}
which manifestly has no singularity until the geometrical focus where
$(R_{m,Q_{2}})_{1}+s_{1}\rightarrow0$. The minus sign comes from the manually
inserted phase shift of the first ray. (All aforementioned quantities should
be expressed in terms of the angles $\theta_{i,1}$ and $\theta_{i,2}\equiv0$
as described in detail in section \ref{sect instruction} below.) The structure
of the first two lines in (\ref{eq Compensated divergence}) is general and is
valid for arbitrary axially symmetric systems. $D_{1}$ is always finite on the
axis, since both the sagittal radius of curvature and the phase difference
$\varphi_{1}-\varphi_{3}$ are proportional to the distance $\rho$.

Equation (\ref{eq Electric field on axis}) is valid even near the focus, since
the diverging terms $D_{1}/Z$ and $(1-z/f)^{-1}$ almost cancel. For
$z\rightarrow f$, however, the divergence of $D_{1}$ itself becomes important,
as the non-paraxial ray 1 becomes axial.

\begin{figure}[t]
\begin{center}
\includegraphics{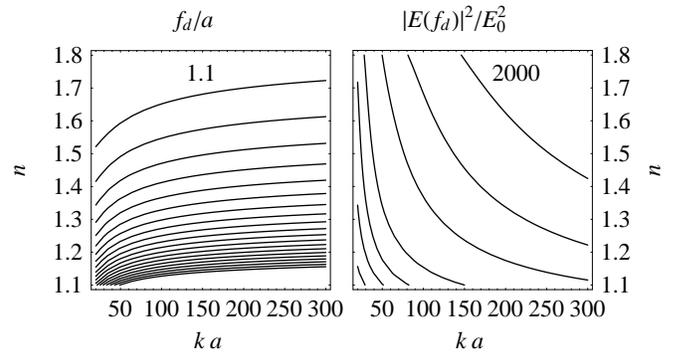}
\end{center}
\par
\vspace{-0.25cm} \caption{Left: Diffraction focus in units of the sphere
radius as a function of $n$ and $k\,a$ (contour lines from top to bottom go
from 1.1 to 3.0 in steps of 0.1). Right: Intensity enhancement at $f_{d}$
(contour lines from bottom left to top right are 20, 50, 100, 200, 500, 1000
and 2000)}%
\label{fig Contouraxis}%
\end{figure}

In figure \ref{fig Contouraxis} we show the position and the value of the
maximum of $\left|  E\right|  ^{2}$ as a function of the refractive index and
the dimensionless product $k\,a$, calculated from (\ref{eq Electric field on
axis}). The $z$-coordinate of this global maximum is denoted with $f_{d}$
(\textit{diffraction focus}) and the intensity there is $\left|
E(f_{d})\right|  ^{2}$. Contrary to the square dependence for the case of an
ideal lens,\cite{Bor2002} even for macroscopic spheres the maximum intensity
turns out to be about proportional to $k\,a$, in agreement with the general
theory.\cite{Kra1999}

The main contribution in (\ref{eq Electric field on axis}) stems from the
Bessoid integral, that is from the term $\propto T_{1}\,D_{1}\,I$. Thus, the
position of the maximum can be estimated from condition (\ref{eq Phase
difference axis}), i.e., $\varphi_{1}-\varphi_{2}\approx3\,\pi/4$. If the
phase difference $\varphi_{1}-\varphi_{2}$ from (\ref{eq Phases sphere 1}) and
(\ref{eq Phases sphere 2}) is expressed as a function of $\theta_{i,1}$,
Taylor expanded and equated to $3\,\pi/4$, then we get in the lowest
non-trivial order of the inverse product $k\,a$:%
\begin{equation}
f_{d}\approx\dfrac{a}{2}\,\dfrac{n}{n-1}\left(  1-\sqrt{\dfrac{3\,\pi
}{4\,k\,a}\,\dfrac{n\,(3-n)-1}{n\,(n-1)}}\,\right)  \!.
\label{eq Diffraction focus}%
\end{equation}
Hence, in the limit of small wavelengths or large spheres the relative
difference between the diffraction and the geometrical focus decreases
proportionally to the inverse square root of $k\,a$. The factor $3\,\pi
/4\approx2.356$ can be replaced by the more exact Bessoid value 2.327 from
(\ref{eq Phase difference axis}). Expression (\ref{eq Diffraction focus})
approximates the position of the maximum within an error of $<5\,\%$ for
$k\,a>100$ and values of the refractive index in the range $1.4<n<1.6$. The
transcendental phase difference condition (\ref{eq Phase difference axis}),
which holds for large angles, naturally has a wider range of applicability.
With very good accuracy the diffraction focus also provides the maximum for
the absolute square of the magnetic field, $\left|  H\right|  ^{2}%
\equiv\mathbf{H}\,\mathbf{H}^{\ast}$, as well as for the $z$-component of the
Poynting vector $\mathbf{S}$. Note that on the axis $\mathbf{H}=H\,\mathbf{e}%
_{y}$ and $\mathbf{S}=S\,\mathbf{e}_{z}$.

\subsection{A protocol for the electromagnetic field calculation behind the
sphere\label{sect instruction}}

For convenience we summarize the sequence of steps that should be used for the
calculation of the field behind a sphere irradiated with linearly polarized
light on the basis of the formulas developed above:

\begin{enumerate}
\item \textit{Finding the rays}: The origin of the coordinate system is in the
center of the sphere. Choose a point $P\equiv(\rho,\varphi,z)$ behind the
sphere and numerically calculate the 3 rays arriving at $P$. These rays are
characterized by the 3 incident angles $\theta_{i,j}$ ($j=1,2,3$), found
numerically from (\ref{eq Cubic like sphere}), and numbered according to
figure \ref{fig Rays}. All other angles follow from Snell's law and from
(\ref{eq alpha beta}), respectively. Outside the cuspoid the rays 2 and 3 are complex.

\item \textit{The geometrical optics solution}: Compute the geometrical optics
solution for the electric field (\ref{eq Electric field components}), which is
the sum of contributions from three rays. The eikonals $\psi_{j}$ are
calculated from (\ref{eq Eikonal sphere}) with (\ref{eq Path}). The
geometrical optics amplitudes are given by the Fresnel transmission components
$T_{j}^{(m)}$ (\ref{eq Transmission components}) and the generalized
divergence factors $1/\sqrt{J_{j}}$ (\ref{eq Divergence sphere}), which follow
from the radii of curvature (\ref{eq R1})--(\ref{eq R4}) and the distances
$s_{j}$ from the sphere to $P$ (\ref{eq Path}). The conventions for the
complex roots shall correctly add up all individual caustic phase shifts.
Ignore the fact that the geometrical field diverges near caustic regions.

\item \textit{Bessoid matching}: Starting from the eikonals $\psi_{j}$,
determine the Bessoid coordinates, i.e., first $Z$, and then $R$ and $\chi$
(\ref{eq R, Z, chi}). Next, compute and correctly order the points of
stationary phase $t_{j}$ (\ref{eq Stationary points}), most conveniently using
trigonometric formulas.\cite{Bro2004} With the generalized divergences
(\ref{eq Divergence sphere}) and the Hessians (\ref{eq Hessian general}) the
Bessoid amplitudes $A_{m}$, $A_{mR}$ and $A_{mZ}$ can be computed from
(\ref{eq Am}) for all orders $m=0,1,2$. The electric field component $E^{(m)}$
associated with the $m$-th order results from the Ansatz (\ref{eq Ansatz
higher-order}). The Bessoid-matched field $\mathbf{E}$ is finally given by
(\ref{eq Electric field matched}). In the case of a scalar plane wave (or
exactly on the axis) only the order $m=0$ contributes. Proceed accordingly for
the magnetic field $\mathbf{H}$, employing different transmission coefficients
in step 2.

\item \textit{The Bessoid integral}: The final solution (\ref{eq Electric
field matched}) contains the Bessoid integral and higher-order Bessoid
integrals as well as their partial derivatives. The Bessoid integral $I(R,Z)$
can be efficiently computed numerically via the differential equation (\ref{eq
Ordinary}). The higher-order Bessoid integrals follow from a recursive
relation (\ref{eq Bessoid recursive}).

\item \textit{Remarks}: The speed limiting bottleneck of this procedure is the
finding of the rays $\theta_{i,j}$ in step 1. The numerical evaluation of the
Bessoid integral is very efficient and in all other steps analytical
expressions are applied. When approaching the caustic (axis, cuspoid),
individual quantities --- the geometrical optics amplitudes --- diverge but
their combinations remain finite. In our calculations we observed perfect
numerical stability up to distances from the caustic of the order of $10^{-5}$
times the sphere radius, which is more than enough for any practical purposes.
\end{enumerate}

\subsection{Comparison with the theory of Mie}

We presented a general way to match geometrical optics solutions with the
Bessoid integrals. It can be applied to any axially symmetric system with the
cuspoid topology of spherical aberration.

\begin{figure}[t]
\begin{center}
\includegraphics{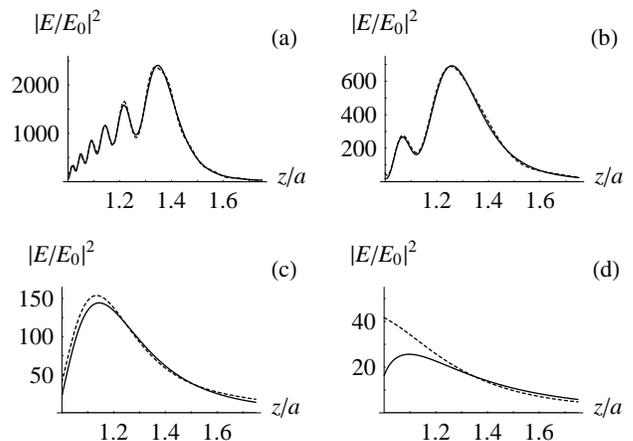}
\end{center}
\par
\vspace{-0.25cm} \caption{$\left|  E/E_{0}\right|  ^{2}$ on the axis. Dashed
lines represent the Mie theory, solid lines are the results of Bessoid
matching. The parameters are as in figure \ref{fig Contourxzyz} and the cases
(a), (b), (c) and (d) correspond to $q=300$, $100$, $30$ and $10$,
respectively. In dimensional units, for $\lambda=0.248\,$$\mu$m this
corresponds to sphere radii of $a\approx12\,$$\mu$m, $4\,$$\mu$m, $1.2\,\mu$m
and $0.4\,\mu$m.}%
\label{fig Comparisonaxis}%
\end{figure}

For the sphere we can compare our approximate results with the theory of
Mie.\cite{Mie1908} A main quantity characterizing the sphere is the
dimensionless Mie parameter $q\equiv k\,a$. Figure \ref{fig Comparisonaxis}
compares the intensity on the axis obtained from the Mie theory with the
Bessoid approximation. The parameters are as in figure \ref{fig Contourxzyz}
and the Mie parameter is $q=300$, $100$, $30$ and $10$.

We see very good agreement down to $q\approx30$ ($a/\lambda\approx4.8$). For
$q=10$ ($a/\lambda\approx1.6$) the asymptotic behavior far from the sphere is
still correct. However, for small $q$ the characteristic scale $a$ is no
longer large compared to the wavelength $\lambda$ and geometrical optics
becomes invalid.

\begin{figure}[t]
\begin{center}
\includegraphics{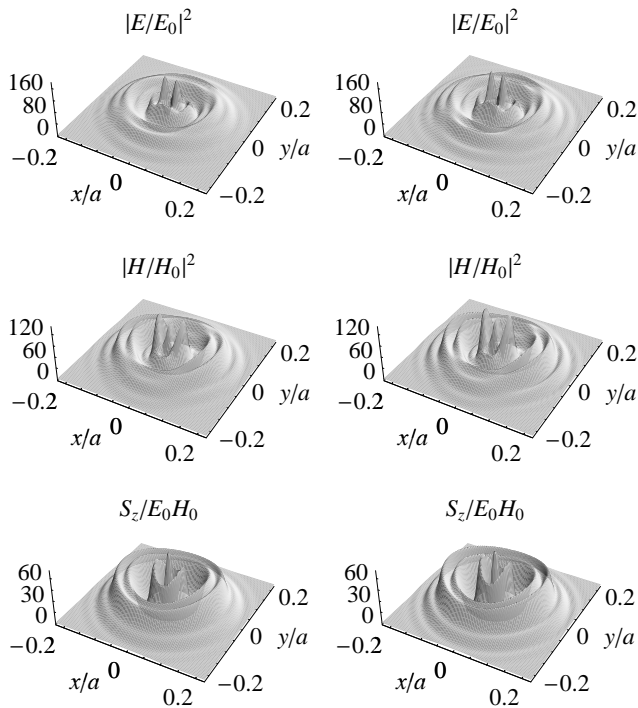}
\end{center}
\par
\vspace{-0.5cm} \caption{Normalized $\left|  E\right|  ^{2}$, $\left|
H\right|  ^{2}$ and $S_{z}$ in the normalized $x$,$y$-plane for $z=1.02\,a$
calculated with Bessoid matching (left) and with the Mie theory (right). The
parameters are the same as in figure \ref{fig Contourxzyz}.}%
\label{fig Comparisonmie}%
\end{figure}

Next, we compare the off-axis electric and magnetic field as well as the
$z$-component of the Poynting vector, $S_{z}$ (figure \ref{fig Comparisonmie}%
). Right behind the sphere ($z=a$) the agreement is not perfect (see figure
\ref{fig Comparisonaxis}), though all qualitative features are preserved.
Sections at $z=1.02\,a$ already show good agreement (figure \ref{fig
Comparisonmie}) and for $z\gtrsim1.05\,a$ the pictures become visually almost
indistinguishable. Discussing the quality of these results, one has to
differentiate between the accuracy of the method and the influence of those
factors which can be taken into account, but were not included into the
current consideration.

The accuracy of the Bessoid matching procedure itself was studied separately
for the case of a spherically aberrated wave incident onto an aperture. The
deviation --- defined as the maximal relative error of the intensity
$(|E_{\text{Bessoid}}|^{2}\!-\!|E_{\text{exact}}|^{2})/|E_{\text{exact}}|^{2}$
--- between the Bessoid matched geometrical optics solution and the
corresponding (exact) Rayleigh--Sommerfeld diffraction integral\cite{Bor2002}
decreases as the aperture increases. If the aperture is large enough, the
deviation is below $10^{-3}$ for spherical abberation strengths and
wavevectors approximately corresponding to the focusing by spheres studied in
figures \ref{fig Comparisonaxis} and \ref{fig Comparisonmie}.

For the sphere, for all investigated Mie parameters $30\leq q\leq300$, the
deviations from the exact Mie solution are about $\pm5\,\%$ in those regions
where the intensity is not very small (including the caustic axis and
cuspoid). A detailed analysis indicates that this deviation originates from
several factors:

(i) Influence of the finite size of the sphere. There exist diffractive
contributions from creeping rays\cite{Kel1962} propagating along the sphere
surface. They can in principle be accounted for at the expense of the
simplicity of the procedure, essentially by considering the interference of
the Bessoid field with these additional rays. This results in oscillations
which can actually be seen in the Mie curve on the right side of the Bessoid
tail in figures \ref{fig Comparisonaxis}a and b.

(ii) Rays entering the sphere undergo multiple interior reflections. Some of
them satisfy resonance conditions, accumulate significant energy inside the
sphere and refract outside. This produces an additional field behind the
sphere, in particular the intensity becomes non-zero in the regions of
geometrical shadow for the directly transmitted rays used in the Bessoid
matching. Here again, one can (in principle) study such multiply reflected
rays separately and add them to the Bessoid field, but their contribution to
the focusing properties of the microspheres is of secondary importance.

(iii) Finally, very close to the sphere surface at distances of the order of a
fraction of $\lambda$, there exist evanescent contributions, which are taken
into account in the Mie theory, but are obviously absent in the Bessoid
matching procedure.

Thus, the quality of the Bessoid matching in the most interesting regions near
caustic surfaces is quite satisfactory. It rectifies the divergences of
geometrical optics, which is asymptotically correct for large $q$ in
non-singular regions of space. Clearly, the procedure has to be extended
whenever contributions from rays other than those 3 used for matching become significant.

The field distribution behind the sphere has a rich fine structure (figure
\ref{fig Comparisonmie}) which our geometrical approach helps to clarify. It
is known that the ring-type field enhancement corresponds to the cuspoid
caustic, having the approximate radial distance\cite{Luk2003}%
\begin{equation}
\rho_{\text{c}}=a\,\dfrac{(4-n^{2})^{3/2}}{3\,\sqrt{3}\,n^{2}}\,.
\end{equation}
Our approach explains the double-peak structure of $\left|  E\right|  ^{2}$
along the direction of polarization. It is related to the axial field
component $E_{z}$ and can be understood in terms of geometrical optics. On the
axis, the $E_{z}$ components from the rays 1 and 3 point into opposite
directions and cancel, having an effective phase difference of $\pi$. Off the
axis, ray 1 underwent a caustic phase shift of $-\pi/2$ when crossing the
axis, which makes the condition for constructive interference: $\varphi
_{1}-\varphi_{3}=3\,\pi/2$. Then, according to (\ref{eq Bessel argument}), the
peak occurs at the radial distance%
\begin{equation}
\rho_{\text{p}}\approx\dfrac{\varphi_{1}-\varphi_{3}}{2\,k\,\sin\beta}%
=\dfrac{3}{8}\,\dfrac{\lambda}{\sin\beta}\,.
\end{equation}
More details on the derivation are given elsewhere\cite{Kof2004} together with
the refined coefficient 0.293 (instead of $3/8$) obtained from the Bessoid
asymptotic (\ref{eq Higher-order Bessoid axis}).

Double-peak structures have been observed in nano-patterning
experiments\cite{Mue2001,Bae2003,Lan2005} and were semi-quantitatively
explained on the basis of the Mie solution.\cite{Luk2002} In an actual
experiment it may depend on the laser pulse parameters and the properties of
the patterned material, whether the Poynting vector or the electric field is
responsible for the patterning process. For small spheres this double peak
effect can be understood using the near field pattern for a scattering
dipole.\cite{Luk2002,Luk2003} The present explanation (for sphere diameters of
a few wavelengths and larger) results in the same orientation of the maxima
and thus these two limiting cases cover almost all range of sphere sizes.
Similar polarization dependence of the field distribution in focal regions can
be used to improve the resolution.\cite{Dor2003}

\section{Conclusions}

We described theoretically arbitrary axially symmetric aberrated focusing and
studied light focusing by microspheres as an example. Following the method of
uniform caustic asymptotics,\cite{Kra1999} we introduced a canonical integral
describing the wave field for the given cuspoid ray topology. This Bessoid
integral appears naturally in the paraxial approximation. In some regions (off
the caustic or exactly on the axis) it reduces to simple analytical
expressions. In other regions we efficiently computed this highly oscillatory
integral via a single ordinary differential equation.

For arbitrary axially symmetric focusing, coordinate and amplitude
transformations match the Bessoid wave field and the solution of geometrical
optics. The caustic divergences of the latter are removed thereby. For
vectorial problems with angularly dependent field components, higher-order
Bessoid integrals are used for the matching procedure. The formulas
significantly simplify on and near the axis. An approximate universal
condition for the diffraction focus can be given in terms of phase
differences. Here, the concept of caustic phase shifts is of main importance.

The central part of the Bessoid integral is essentially a Bessel
beam\cite{McG2005} with a variable cross section due to the variable angle of
the non-paraxial rays. Its local diameter is always smaller than in the focus
of an ideal lens with the same numerical aperture. Besides, the largest
possible apertures can be physically realized, which is hardly possible with
lenses. All this is achieved at the expense of longitudinal confinement.

As an example the focusing of a linearly polarized plane wave by a transparent
sphere is studied in detail. We calculate the geometrical optics eikonals and
divergences, incorporate Fresnel transmission coefficients and perform Bessoid
matching. Using the general theory, simple expressions for the light field on
the axis and for the diffraction focus are derived. The two strong maxima in
the intensity observed immediately behind the sphere can be explained as well.

Finally, the results of the Bessoid matching procedure are compared with the
Mie theory. The agreement is good for Mie parameters $k\,a>30$. Near the
sphere the correspondence is worse due to unaccounted evanescent contributions.

The developed formulas can be directly applied in other areas of physics where
non-paraxial axially symmetric focusing is of importance, e.g., acoustics,
semiclassical quantum mechanics,\cite{Pet1997} flat superlenses based on
left-handed materials,\cite{Par2003} radio wave propagation, scattering
theory,\cite{Con1981} chiral conical diffraction\cite{Ber2006} etc.

Concluding, let us briefly enumerate several possibilities to extend and
refine the developed formalism. Weak absorption can be incorporated easily,
for it just changes the amplitudes along the rays and the transmission
coefficients. Strong absorption additionally modifies Snell's law of
refraction, still preserving the axial symmetry. One can consider incoming
radially or azimuthally polarized beams, which are known to produce better
resolution than linear polarization.\cite{Dor2003} The diffraction of light
from regions beyond the sphere radius can be incorporated by considering the
interference of the Bessoid field with creeping rays.\cite{Kel1962} For other
geometries, in particular finite apertures with sharp boundaries, edge rays or
the Rubinowitz representation,\cite{Bor2002} or an approach based on
catastrophe theory\cite{Nye2005} have to be used. Such corrections become
relevant, for example, for the ray structure and the field distribution
immediately behind spheres with a refractive index $n<\sqrt{2}$. Finally, one
can calculate the interference of the diffracted light with the original
incident wave or the interference of the light refracted by several spheres or
arrays of spheres. The latter yields interesting secondary patterns
\cite{Bae2002} related to the so called Talbot effect.\cite{Ber2001}

\subsection*{Acknowledgments}

The authors thank D.~B\"{a}uerle (Johannes Kepler University, Linz) for many
stimulating discussions on microsphere patterning experiments, which initiated
this study, and for his continuous support of this work. The authors also
thank B.~Luk'yanchuk and Z.~B.~Wang (both at the Data Storage Institute,
Singapore) for their Mie program and discussions on Mie calculations.
J.~K.\ appreciates helpful conversations with G.~Langer (Johannes Kepler
University, Linz). N.~A.\ thanks V.~Palamodov (Tel Aviv University) for
illuminating mathematical suggestions. Financial support was provided by the
FWF (Austrian Science Fund) under Contract No.\ P16133--N08. N.~A.\ also
thanks the Christian Doppler Laboratory of Surface Optics (Johannes Kepler
University, Linz).

\appendix

\section{A near axis approximation for the Bessoid integral\label{Appendix
Near Axis Bessoid}}

We make the substitution $w\equiv\rho_{1}^{2}$ in (\ref{eq Bessoid polar}):%
\begin{equation}
I(R,Z)=\dfrac{1}{2}\,\int\nolimits_{0}^{\infty}J_{0}(R\sqrt{w})\;\text{e}%
^{-\,\text{i}\,\left(  Z\,\tfrac{w}{2}\,+\,\tfrac{w^{2}}{4}\right)  }%
\text{d}w\,. \label{eq Bessoid substituted}%
\end{equation}
Near the axis (small $R$) the Bessel function is slowly varying compared with
the exponent. The integral will have significant contribution only from the
region in which the exponent's phase is stationary, i.e., regions near $w=-Z$.
We consider the most interesting caustic part of the axis for which $Z\leq0$.
In a lowest order approximation the Bessel function is considered as constant
near the stationary point $-Z$ and can be pulled out of the integral. The
phase can be written as a complete quadratic form. With the full square of
$v\equiv(w+Z)/2$:%
\begin{equation}
I(R,Z)\approx J_{0}(R\sqrt{-Z})\;\text{e}^{\,\text{i}\,\tfrac{Z^{2}}{4}}%
\int\nolimits_{Z/2}^{\infty}\text{e}^{-\,\text{i}\,v^{2}}\text{d}v\,.
\label{eq Bessoid quadratic}%
\end{equation}
The remaining integral can be expressed in terms of the complementary error
function\cite{Abr1993} erfc (of complex argument) and hence we arrive at
(\ref{eq Near axis}).

For $Z>0$ the point $w=0$ should be taken as a stationary edge point of the
integration.\cite{Foc1956} And the near axis approximation (\ref{eq Near
axis}) remains valid as long as the Bessel function is set to $J_{0}(0)=1$.

\section{An ordinary differential equation for the Bessoid
integral\label{Appendix Differential}}

We derive the paraxial Helmholtz equation%
\begin{equation}
I_{RR}+\dfrac{1}{R}\,I_{R}+2\,\text{i}\,I_{Z}=0\,,
\label{eq Differential Bessoid 1}%
\end{equation}
as well as the following \textit{ordinary differential equation} for the
Bessoid integral:%
\begin{equation}
I_{RRR}+\dfrac{1}{R}\,I_{RR}-\left(  \dfrac{1}{R^{2}}+Z\right)  I_{R}%
+\text{i}\,R\,I=0\,. \label{eq Differential Bessoid 2}%
\end{equation}
Indices denote partial derivatives. Both equations can be rewritten in the
compact form%
\begin{align}
L+2\,\text{i}\,I_{Z}  &  =0\,,\\
L_{R}-Z\,I_{R}+\text{i}\,R\,I  &  =0\,,
\end{align}
where $L$ is the radial Laplacian%
\begin{equation}
L\equiv I_{RR}+\dfrac{1}{R}\,I_{R}\,.
\end{equation}
We begin with the proof of (\ref{eq Differential Bessoid 1}) and state that we
may differentiate under the integral sign, since the partial derivatives of
the integrand exist and are continuous functions. Starting from the Bessoid
integral in the polar representation (\ref{eq Bessoid polar}), its integrand
can be written as%
\begin{equation}
G\equiv\rho_{1}\,J_{0}(R\,\rho_{1})\,E\,,
\end{equation}
with the abbreviation%
\begin{equation}
E\equiv\text{e}^{-\,\text{i}\,\left(  Z\,\tfrac{\rho_{1}^{2}}{2}%
\,+\,\tfrac{\rho_{1}^{4}}{4}\right)  }.
\end{equation}
The (multiple) partial derivatives are%
\begin{align}
G_{R}  &  =-\rho_{1}^{2}\,J_{1}(R\,\rho_{1})\,E\,,\label{eq GR}\\
G_{RR}  &  =-\dfrac{\rho_{1}^{3}}{2}\,[J_{0}(R\,\rho_{1})-J_{2}(R\,\rho
_{1})]\,E\,,\label{eq GRR}\\
G_{Z}  &  =-\text{i}\,\dfrac{\rho_{1}^{3}}{2}\,J_{0}(R\,\rho_{1})\,E\,.
\end{align}
Here we used the derivative formula for Bessel functions\cite{Abr1993}%
\begin{equation}
\dfrac{\text{d}}{\text{d}t}\,J_{m}(t)=\dfrac{J_{m-1}(t)-J_{m+1}(t)}{2}
\label{eq Bessel derivative}%
\end{equation}
with $m=0$ to obtain (\ref{eq GR}) and $m=1$ for (\ref{eq GRR}). Note that
$J_{-1}(t)=-J_{1}(t)$. Applying the recurrence relation for Bessel
functions\cite{Abr1993}%
\begin{equation}
J_{m+1}(t)=-J_{m-1}(t)+\dfrac{2\,m}{t}\,J_{m}(t)\,,
\label{eq Bessel recurrence}%
\end{equation}
one can eliminate $J_{2}$ from (\ref{eq GRR}). And then it is enough to notice
and verify that%
\begin{equation}
G_{RR}+\dfrac{1}{R}\,G_{R}+2\,\text{i}\,G_{Z}=0\,.
\end{equation}
This proves equation (\ref{eq Differential Bessoid 1}).

For the proof of (\ref{eq Differential Bessoid 2}), we need to note that its
left hand side can be expressed as the integral of a partial derivative%
\begin{equation}
H\equiv%
{\displaystyle\int\nolimits_{0}^{\infty}}
\dfrac{\partial}{\partial\rho_{1}}\!\left[  \,\text{i}\,\rho_{1}%
\,J_{1}(R\,\rho_{1})\;\text{e}^{-\,\text{i}\,\left(  Z\,\tfrac{\rho_{1}^{2}%
}{2}\,+\,\tfrac{\rho_{1}^{4}}{4}\right)  }\right]  \!\text{d}\rho_{1}.
\label{eq Definite integral}%
\end{equation}
With the help of (\ref{eq Bessel derivative}) and (\ref{eq Bessel recurrence})
both the left hand side of (\ref{eq Differential Bessoid 2}) and $H$ become%
\begin{equation}%
{\displaystyle\int\nolimits_{0}^{\infty}}
[\,\text{i}\,R\,\rho_{1}\,J_{0}(R\,\rho_{1})+\rho_{1}^{2}\,(Z+\rho_{1}%
^{2})\,J_{1}(R\,\rho_{1})]\,E\,\text{d}\rho_{1}.
\end{equation}
Thus, in order to prove (\ref{eq Differential Bessoid 2}), it is enough to
show that $H=0$. This follows from the Newton-Leibniz formula applied to the
(definite) integral (\ref{eq Definite integral}):%
\begin{equation}
H=\left[  \,\text{i}\,\rho_{1}\,J_{1}(R\,\rho_{1})\;\text{e}^{-\,\text{i}%
\,\left(  Z\,\tfrac{\rho_{1}^{2}}{2}\,+\,\tfrac{\rho_{1}^{4}}{4}\right)
}\right]  _{0}^{\infty}=0\,.
\end{equation}
The lower bound at $0$ vanishes for obvious reasons. For the upper bound at
$\infty$ one assumes an \textit{infinitely small} imaginary part in front of
the fourth order term in the exponent: $\rho_{1}^{4}\rightarrow(1-\,$%
i$\,\varepsilon)\,\rho_{1}^{4}$ with $\varepsilon>0$. This completes the proof
of the differential equations (\ref{eq Differential Bessoid 1}) and (\ref{eq
Differential Bessoid 2}) for the Bessoid integral.

\section{The near axis Bessoid coordinates\label{Appendix Near axis
coordinates}}

\begin{figure}[t]
\begin{center}
\includegraphics{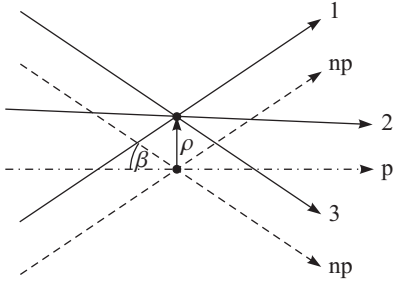}
\end{center}
\par
\vspace{-0.25cm}\caption{Near the axis, the phases of the rays 1 and 3 differ
from the phase of the non-paraxial rays np (crossing the axis at the same $z$)
by $\pm\,k\,\rho\,\sin\beta$, whereas the phases of ray 2 and the (par)axial
ray are the same in first order. The non-paraxial ray and ray 3 cross the axis
at an angle $\beta>0$.}%
\label{fig Nearaxis}%
\end{figure}

Near the axis the phases of the rays can be Taylor expanded. From figure
\ref{fig Nearaxis} one infers that up to the first order in $\rho$ the phases
can be written as%
\begin{align}
\varphi_{1}  &  \approx\varphi_{\text{np}}+k\,\rho\,\sin\beta\,,\nonumber\\
\varphi_{2}  &  \approx\varphi_{\text{p}}\,,\label{eq Phases Taylor}\\
\varphi_{3}  &  \approx\varphi_{\text{np}}-k\,\rho\,\sin\beta\,.\nonumber
\end{align}
Here $\varphi_{\text{np}}$ and $\varphi_{\text{p}}$ denote the phases of the
non-paraxial rays and the (par)axial ray (with $\rho=0$) and $\beta>0$ is the
angle of ray 3 with the axis.

We insert these phases into the exact expressions for $R$ and $Z$ in (\ref{eq
R, Z, chi}), Taylor expand the result with respect to $\rho$ and resubstitute
$\varphi_{\text{np}}\approx(\varphi_{1}+\varphi_{3})/2$, $\varphi_{\text{p}%
}\approx\varphi_{2}$ and $k\,\rho\,\sin\beta\approx(\varphi_{1}-\varphi
_{3})/2$ from (\ref{eq Phases Taylor}). This yields (\ref{eq Near axis R}) and
(\ref{eq Near axis Z}).

\section{The on axis field\label{Appendix On axis}}

Here we derive a simple on axis expression for the Bessoid-matched field $U$
(\ref{eq Ansatz}), namely equation (\ref{eq On axis field}).

On the axis and inside the cusp $\rho=0$ ($R=0$) and $z<0$ ($Z<0$). The
stationary points, given by (\ref{eq Stationary points}), are%
\begin{equation}
t_{1}=-\sqrt{-Z}\,,\quad t_{2}=0\,,\quad t_{3}=-t_{1}\,. \label{eq ts on axis}%
\end{equation}
Then, the amplitude $A$ in (\ref{eq A, AR, AZ}) simplifies to%
\begin{equation}
A=U_{0,2}\,\dfrac{\sqrt{H_{2}}}{\sqrt{J_{2}}}\,,
\end{equation}
because on the axis the ratios $\sqrt{H_{1,3}}/\sqrt{J_{1,3}}$ are both finite
and the corresponding other terms disappear upon multiplication with $t_{2}%
=0$. Due to the restriction to the lit region ($Z<0$), all rays are real and
equation (\ref{eq Hessian lit}) holds. By virtue of (\ref{eq Det H})
$\det\mathbf{H}_{2}=Z^{2}$, and due to (\ref{eq Sign H}) sign$\,\mathbf{H}%
_{2}=2$, one finds%
\begin{equation}
\sqrt{H_{2}}=\text{i}\,Z\,, \label{eq Sqrt H2}%
\end{equation}
and thus:%
\begin{equation}
A=\text{i}\,\dfrac{U_{0,2}\,Z}{\sqrt{J_{2}}}\,.
\end{equation}
This approximation for the amplitude $A$ is valid up to the focus ($Z=0$). As
ray 2 converges like the inverse distance from the focus, $\sqrt{J_{2}}$ is
proportional to $Z$.

The amplitude $A_{R}$ in (\ref{eq A, AR, AZ}) vanishes due to $t_{3}=-t_{1}$
and the fact that%
\begin{equation}
\dfrac{\sqrt{H_{1}}}{\sqrt{J_{1}}}=\dfrac{\sqrt{H_{3}}}{\sqrt{J_{3}}}\,,
\label{eq Ray 1 and 3}%
\end{equation}
which means that rays 1 and 3 have equal amplitudes and that the caustic phase
shifts are in accordance with the signature of the Hessian. Consequently,%
\begin{equation}
A_{R}=0\,.
\end{equation}

With (\ref{eq ts on axis}) and (\ref{eq Ray 1 and 3}) the amplitude $A_{Z}$
reads%
\begin{equation}
A_{Z}=\dfrac{2}{Z}\left(  U_{0,1}\,\dfrac{\sqrt{H_{1}}}{\sqrt{J_{1}}}%
-U_{0,2}\,\dfrac{\sqrt{H_{2}}}{\sqrt{J_{2}}}\right)  \!.
\end{equation}
The first term is non-trivial. Both $\sqrt{H_{1}}$ and $\sqrt{J_{1}}$ are zero
on the axis, but their ratio is finite and well defined. Indeed, the Taylor
expansion of Cardan's solution $t_{1}$ in its trigonometric
representation\cite{Bro2004} yields in the first order in $R$:%
\begin{equation}
t_{1}=-\sqrt{-Z}+\dfrac{R}{2\,Z}\,.
\end{equation}
Therefore, again in first order in $R$: $\det\mathbf{H}_{1}=2\,R\,\sqrt{-Z}$.
Due to sign$\,\mathbf{H}_{1}=-2$ we obtain%
\begin{equation}
\sqrt{H_{1}}=\text{i}\,\sqrt{2\,R\,\sqrt{-Z}}\,,
\end{equation}
and with (\ref{eq Sqrt H2}):%
\begin{equation}
A_{Z}=2\,\text{i}\,\dfrac{U_{0,1}\sqrt{2\,R\,\sqrt{-Z}}}{Z\,\sqrt{J_{1}}%
}-2\,\text{i}\,\dfrac{U_{0,2}}{\sqrt{J_{2}}}\,.
\end{equation}
This approximation for $A_{Z}$ holds for small values of $R$. It is finite,
since $\sqrt{J_{1}}$ approaches zero as $\sqrt{R}$ for $R\rightarrow0$.

For the final representation of the field $U$, we can substitute the near axis
expression for $R\sqrt{-Z}$ (\ref{eq Bessel argument}) into $A_{Z}$. On the
axis the phase coordinate becomes $\chi=\varphi_{2}$, which results from
substituting $\varphi_{1}=\varphi_{3}$ and $Z=-2\,\sqrt{\varphi_{1}%
-\varphi_{2}}$ into the corresponding expression in (\ref{eq R, Z, chi}). This
leads to%
\begin{align}
U  &  =\left(  A\,I+\dfrac{1}{\text{i}}\,A_{R}\,I_{R}+\dfrac{1}{\text{i}%
}\,A_{Z}\,I_{Z}\right)  \text{e}^{\,\text{i}\,\chi}\\
&  =\left(  \dfrac{U_{0,2}}{\sqrt{J_{2}}}\,(\text{i}\,Z\,I-2\,I_{Z}%
)+\dfrac{2\,U_{0,1}\,\sqrt{2\,k\,\rho\,\sin\beta}}{Z\,\sqrt{J_{1}}}%
\,I_{Z}\right)  \text{e}^{\,\text{i}\,\varphi_{2}}.\nonumber
\end{align}
Using the linear relationship between the Bessoid integral and its
$Z$-derivative (\ref{eq Condition 3}),%
\begin{equation}
\text{i}\,Z\,I-2\,I_{Z}=1\,,
\end{equation}
we end up with equation (\ref{eq On axis field}).

\section{Higher-order Bessoid integrals\label{Appendix Higher-order}}

\textit{Higher-order Bessoid integrals} (\ref{eq Higher-order Bessoids})
appear naturally, if one expands an arbitrary initial field amplitude on the
aperture in a Fourier series:%
\begin{equation}
U_{0}(\rho_{1},\varphi_{1})=%
{\displaystyle\sum\limits_{m=0}^{\infty}}
\,[\,a_{m}(\rho_{1})\cos(m\,\varphi_{1})+b_{m}(\rho_{1})\sin(m\,\varphi
_{1})\,]\,.
\end{equation}
The form of the coefficients $a_{m}$ and $b_{m}$ can be seen from a
two-dimensional Taylor expansion in Cartesian coordinates around the point
$(0,0)$, rewritten into polar coordinates:%
\begin{equation}
U_{0}(\rho_{1},\varphi_{1})=%
{\displaystyle\sum\limits_{m=0}^{\infty}}
\,%
{\displaystyle\sum\limits_{n=0}^{m}}
\,c_{mn}\,\rho_{1}^{m}\cos^{m-n}\!\varphi_{1}\sin^{n}\!\varphi_{1}\,,
\end{equation}
with%
\begin{equation}
c_{mn}\equiv\dfrac{1}{m!}\,\binom{m}{n}\left.  \dfrac{\partial^{m}U_{0}%
(x_{1}^{\prime},y_{1}^{\prime})}{\partial x_{1}^{\prime m-n}\partial
y_{1}^{\prime n}}\right|  _{x_{1}^{\prime}=0,\,y_{1}^{\prime}=0}\!.
\end{equation}
Thus, $\rho_{1}^{m}$ is the lowest possible power of $\rho_{1}$ which can be
found in the term with $\exp($i$\,m\,\varphi_{1})$. An additional $\rho_{1}$
comes from the transformation from Cartesian to polar coordinates.

If we define the functions%
\begin{equation}
\widetilde{I}_{m}\equiv I_{m}\,\text{e}^{\,\text{i}\,m\,\varphi}\,,
\end{equation}
we find that they satisfy the paraxial Helmholtz equation%
\begin{equation}
2\,\text{i}\,\widetilde{I}_{m,Z}+\widetilde{I}_{m,RR}+\dfrac{1}{R}%
\,\widetilde{I}_{m,R}+\dfrac{1}{R^{2}}\,\widetilde{I}_{m,\varphi\varphi}=0\,,
\end{equation}
where $\widetilde{I}_{m,\varphi\varphi}=-m^{2}\,\widetilde{I}_{m}$.

Due to (\ref{eq Bessel derivative}) and (\ref{eq Bessel recurrence}), one can
write the identity%
\begin{align}
\rho_{1}^{m+2}\,J_{m+1}(R\,\rho_{1})  &  =-\dfrac{\partial}{\partial R}%
[\rho_{1}^{m+1}\,J_{m}(R\,\rho_{1})]\nonumber\\
&  \quad\quad+\dfrac{m}{R}\,\rho_{1}^{m+1}\,J_{m}(R\,\rho_{1})\,.
\end{align}
Hence, the recursive relation for the Bessoid integrals (\ref{eq Bessoid
recursive}) follows:%
\begin{equation}
I_{m+1}=-I_{m,R}+\dfrac{m}{R}\,I_{m}\,, \label{eq Bessoid recursive1}%
\end{equation}
i.e., $I_{1}=-I_{0,R}\equiv-I_{R}$, $I_{2}=I_{RR}-I_{R}/R$, etc. Using
(\ref{eq Bessoid recursive1}) and (\ref{eq Bessel recurrence}) as well as
expression (\ref{eq Near axis}) for $I$, one obtains%
\begin{align}
I_{m}(R,Z)  &  \approx\dfrac{\sqrt{\pi}\,(-Z)^{m/2}}{2}\,J_{m}(R\sqrt
{-Z})\nonumber\\
&  \quad\quad\times\text{e}^{\,\text{i}\,\tfrac{Z^{2}-\pi}{4}}%
\,\operatorname{erf}\!\text{c\negthinspace}\left(  \dfrac{Z}{2}\;\text{e}%
^{\,\text{i}\,\tfrac{\pi}{4}}\right)  \text{\negthinspace},
\label{eq Higher-order Bessoid axis}%
\end{align}
which is the analytic near axis expression for the higher-order Bessoid integrals.

While the coordinates and phases ($R$, $Z$, $\chi$) remain unchanged, the
derivation of the higher-order amplitudes ($A_{m}$, $A_{mR}$, $A_{mZ}$)
requires some insight for $m\geq2$. Let us briefly consider the case $m=2$.
For the matching procedure we need the asymptotic behavior of $I_{2}%
=I_{RR}-I_{R}/R$ far from the caustic regions where $R\gg1$ and where it is
dominated by the term $I_{RR}$. Note that for the matching procedure we need
exactly this asymptotic representation and in the non-caustic regions only.
Thus --- although we need the second-order Bessoid integral $I_{2}$ on and
near the axis, where it vanishes --- we shall use its asymptotic stationary
phase expressions far from the axis for the derivation of the amplitudes. In
this region it is equivalent to the asymptotic of $I_{RR}$.

In fact, we may generalize this statement to arbitrary order. Due to (\ref{eq
Bessoid recursive1}) the leading term in the stationary phase calculation is
always%
\begin{equation}
I_{m}\rightarrow\left(  -\dfrac{\partial}{\partial R}\right)  ^{\!m}\!I\,.
\end{equation}
Therefore, the equations for the amplitudes (\ref{eq Matching amplitudes})
become%
\begin{equation}
\dfrac{U_{0,j}^{(m)}}{\sqrt{J_{j}}}=(\text{i}\,t_{j})^{m}\,\dfrac{A_{m}%
-t_{j}\,A_{mR}-\frac{1}{2}\,t_{j}^{2}\,A_{mZ}}{\sqrt{H_{j}}}\,,
\label{eq Matching amplitudes higher-order}%
\end{equation}
which can be seen from the Bessoid integral's Cartesian representation with
the phase (\ref{eq Phase Bessoid}):%
\begin{equation}
\left(  -\dfrac{\partial}{\partial R}\right)  ^{\!m}\!I\rightarrow\left(
-\text{i}\,\dfrac{\partial\phi}{\partial R}\right)  ^{\!m}I=(\text{i}%
\,x_{1})^{m}I\,.
\end{equation}
The equations (\ref{eq Matching amplitudes higher-order}) have the same form
as (\ref{eq Matching amplitudes}) except an additional factor $($%
i$\,t_{j})^{m}$ on the left hand side, proving (\ref{eq Am}).

\section{Wavefront radii of curvature for the refraction on a
sphere\label{Appendix Radii}}

\begin{figure}[t]
\begin{center}
\includegraphics{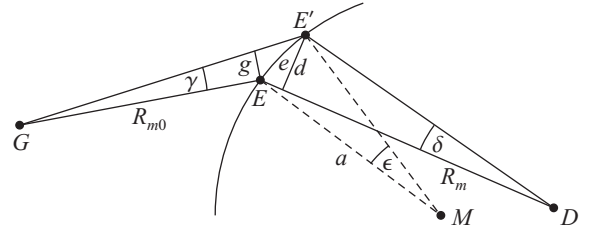}
\end{center}
\par
\vspace{-0.25cm}\caption{Meridional cross section for the determination of the
meridional radius of curvature, $R_{m}$. $M$ is the center of the sphere}%
\label{fig Radius1}%
\end{figure}

Let us consider a point source $G$ and start with the derivation of the
meridional radius of curvature (figure \ref{fig Radius1}).

The initial radius of curvature is $R_{m0}\equiv\overline{GE}$, the one after
refraction is $R_{m}\equiv\overline{ED}$. The infinitesimally neighbored beam
($\gamma\ll1)$ which is refracted in $E^{\prime}$ (the angles of incidence and
transmission in $E$ are $\theta_{i}$ and $\theta_{t}$, in $E^{\prime}$ they be
denoted $\theta_{i}^{\prime}$ and $\theta_{t}^{\prime}$) also propagates to
$D$. The normals onto $R_{m0}$ through $E$ and onto $R_{m}$ through
$E^{\prime}$ are $g$ and $d$, respectively. In the necessary order the length
of the arc $EE^{\prime}$ can be approximated by the distance $e\approx
\overline{EE^{\prime}}$. As all angles are small:\ $g=\gamma\,R_{m0}$,
$d=\delta\,R_{m}$, and $e=\epsilon\,a$. On the other hand, we find from the
infinitesimal triangles: $g=e\cos\theta_{i}$, $d=e\cos\theta_{t}$. This leads
to%
\begin{equation}
R_{m}=-\dfrac{\gamma\,R_{m0}\cos\theta_{t}}{\delta\cos\theta_{i}}\,,
\label{eq R1 first}%
\end{equation}
where we have introduced a minus sign because the wave is converging after the
refraction. The remaining problem is the angle $\delta$ in the denominator. To
find $\delta$ we write the relations between angles and primed angles:%
\begin{equation}
\theta_{i}^{\prime}=\theta_{i}+\gamma+\epsilon\,,\quad\theta_{t}^{\prime
}=\theta_{t}-\delta+\epsilon\,. \label{eq delta}%
\end{equation}
With Snell's law%
\begin{equation}
\theta_{t}-\theta_{t}^{\prime}=\arcsin\dfrac{\sin\theta_{i}}{n}-\arcsin
\dfrac{\sin(\theta_{i}+\gamma+\epsilon)}{n}\,,
\end{equation}
a first order Taylor expansion in $(\gamma+\epsilon)$ yields%
\begin{equation}
\theta_{t}-\theta_{t}^{\prime}=-\dfrac{(\gamma+\epsilon)\cos\theta_{i}}%
{n\cos\theta_{t}}\,, \label{eq theta diff}%
\end{equation}
We express $\delta$ from (\ref{eq delta}), substitute it into (\ref{eq R1
first}) and finally obtain the meridional radius of curvature (\ref{eq Rm})%
\begin{equation}
R_{m}=\dfrac{n\,a\,R_{m0}\cos^{2}\theta_{t}}{a\cos^{2}\theta_{i}+R_{m0}%
\,(\cos\theta_{i}-n\cos\theta_{t})}\,.
\end{equation}

For the sagittal radius of curvature we consider figure \ref{fig
Radius2}.\begin{figure}[t]
\begin{center}
\includegraphics{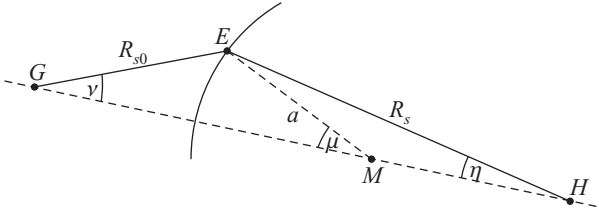}
\end{center}
\par
\vspace{-0.25cm}\caption{Meridional cross section for the determination of the
sagittal radius of curvature, $R_{s}$}%
\label{fig Radius2}%
\end{figure}

A ray which emerged from $G$ is refracted in $E$ (incident angle $\theta_{i}$
and transmitted angle $\theta_{t}$). The distance from $E$ to the intersection
$H$ of the ray with the line passing through $G$ and the sphere center $M$ is
the sagittal radius of curvature, as a neighbored ray, emerging from $G$ and
hitting the sphere not in $E$ but infinitesimally shifted perpendicular to the
meridional plane, will also propagate to $H$ due to symmetry around the line
$GM$. We have $R_{s0}\equiv\overline{GE}$ and $R_{s}\equiv\overline{EH}$. The
tangent theorem states (all angles are in general large now)%
\begin{equation}
\tan\dfrac{\nu-\mu}{2}=\dfrac{a-R_{s0}}{a+R_{s0}}\,\cot\dfrac{\pi-\theta_{i}%
}{2}\,,
\end{equation}
where $\pi-\theta_{i}$ is just the third angle in the triangle $GME$. Due to
$\nu+\mu=\theta_{i}$ we find%
\begin{equation}
\mu=\dfrac{\theta_{i}}{2}-\arctan\!\left(  \dfrac{a-R_{s0}}{a+R_{s0}}%
\,\cot\dfrac{\pi-\theta_{i}}{2}\right)  \!.
\end{equation}
In the triangle $MHE$ the sine theorem reads%
\begin{equation}
-\dfrac{R_{s}}{a}=\dfrac{\sin(\pi-\mu)}{\sin\eta}\,,
\end{equation}
where we have again introduced a minus sign due to the convergence of the
refracted wave. Trigonometric transformations finally give the sagittal radius
of curvature (\ref{eq Rs})%
\begin{equation}
R_{s}=\dfrac{n\,a\,R_{s0}}{a+R_{s0}\,(\cos\theta_{i}-n\cos\theta_{t})}\,.
\end{equation}


\begin{thebibliography}{99}
\bibitem{Time}In the present work all plots of the Bessoid integral (and its
derivatives) contain $101\!\times\!101$ mesh points, computed with the
software package Mathematica 5 (Wolfram Research). Direct numerical
integration of (\ref{eq Bessoid polar}) takes more than one hour on a modern
personal computer. Integration along a line in the complex plane decreases the
amount of time by approximately a factor of 3. Solving the ordinary
differential equation (\ref{eq Ordinary}), however, lasts only a few seconds.
\end{thebibliography}
\end{document}